\documentclass[showpacs,preprintnumbers,amsmath,amssymb]{revtex4}

\usepackage{graphicx}
\usepackage{dcolumn}
\usepackage{bm}

\begin{document}

\preprint{}

\title{The relaxation dynamics of a supercooled liquid confined by rough walls}
\author{Peter Scheidler$^1$}
\author{Walter Kob$^2$}
\author{Kurt Binder$^1$}
\affiliation{$^1$
Institut f\"ur Physik, Johannes Gutenberg-Universit\"at,
55099 Mainz, Germany\\
$^2$ Laboratoire des Verres, Universit\'e Montpellier II,
34095 Montpellier, France
}
\thanks{Author to whom correspondence should be send to}
\email{walter.kob@ldv.univ-montp2.fr}

\date{\today}

\begin{abstract}
We present the results of molecular dynamics computer simulations of a
binary Lennard-Jones liquid confined between two parallel rough walls.
These walls are realized by frozen amorphous configurations of the
same liquid and therefore the structural properties of the confined
fluid are identical to the ones of the bulk system. Hence this setup allows
us to study how the relaxation dynamics is affected by the pure effect
of confinement, i.e. if structural changes are completely avoided. We
find that the local relaxation dynamics is a strong function of $z$,
the distance of the particles from the wall, and that close to the
surface the typical relaxation times are orders of magnitude larger
than the ones in the bulk. Because of the cooperative nature of the
particle dynamics, the slow dynamics also affects the dynamics of the
particles for large values of $z$. Using various empirical laws, we are
able to parameterize accurately the $z-$dependence of the generalized
incoherent intermediate scattering function $F_{\rm s}(q,z,t)$ and also
the spatial dependence of structural relaxation times. These laws allow
us to determine various dynamical length scales and we find that their
temperature dependence is compatible with an Arrhenius law.  Furthermore,
we find that at low temperatures time and space dependent correlation
function fulfill a generalized factorization property similar to the one
predicted by mode-coupling theory for bulk systems. For thin films and/or
at sufficiently low temperatures, we find that the relaxation dynamics
is influenced by the two walls in a strongly non-linear way in that the
slowing down is much stronger than the one expected from the presence of
only one confining wall. Finally we study the average dynamics of all
liquid particles and find that the data can be described very well by a
superposition of two relaxation processes that have clearly separated time
scales. Since this is in contrast with the result of our analysis of the
local dynamics, we argue that a correct interpretation of experimental
data can be rather difficult.

\end{abstract}


\maketitle

\section{Introduction}
\label{sec1}

The origin for the dramatic slowing down of the relaxation dynamics
of a liquid close to the glass transition is still a matter of
debate~\cite{idmrcs4}. Different theoretical approaches to describe
the dynamic transition from the liquid to the glassy state have
been presented so far. Some concentrate on thermodynamic aspects and
propose a first or second order phase transition (e.g. free volume
theory~\cite{turnbull58} or entropy models~\cite{adam65,gibbs58}),
while others, such as mode-coupling theory~\cite{mct}, focus on the
microscopic dynamics that shows non-linear feedback effects which are
responsible for the slow dynamics.

Many of these theories are using in some form the concept of
cooperative motion within a liquid. As already mentioned by
Kauzmann~\cite{kauzmann48} it is reasonable to assume that in a
liquid neighboring particles cannot move independently of each other,
in particular at low temperatures. Instead one must expect that the
motion of two particles that are separated by a distance that is short is
cooperative, i.e. there exist cooperatively rearranging regions (CRR's)
within the liquid. However, how this cooperative motion is realized
microscopically is not that clear. E.g. one can discuss the motion of the
particles by means of the so called cage effect~\cite{mct,sjogren80}:
A liquid particle is surrounded by its nearest neighbors, which form
a cage. At low temperatures this cage is quite rigid and therefore at
short and intermediate times a particle will mainly rattle around within
this cage. For longer times the cage will open up and allow the particle
to escape.  With decreasing temperature this opening will take more and
more time, and hence the relaxation dynamics of the particles will slow
down. The cage effect already includes the concept of cooperativity
since other particles have to move as well before the cage breaks up
and the trapped particle can escape.

Under the assumption that there is some kind of cooperativity within the
liquid the next question is whether the size of such CRR's does indeed
increase with decreasing temperature, as proposed in several models for
the glass transition~\cite{adam65,donth01}. Usual thermodynamic second
order phase transitions, who also show a strong slowing down of
the dynamics, are accompanied by the divergence of some static length
scale. While in spin glasses higher order correlation functions do show a
divergence~\cite{rieger95}, no diverging {\it static} length scale has been found
so far in structural glasses. The analysis of two point correlations
such as the structure factors or radial distribution functions shows
that the length scales involved are hardly growing with decreasing
temperature. In particular, the structure of a glass is almost
identical to the one of a liquid. On the other hand there has been
evidence for the existence of {\it dynamic} length scales which grow much
faster, although evidence for a divergence at some finite critical
temperature has been given only very recently~\cite{berthier03}.
E.g. computer simulations~\cite{donati98,doliwa00} and measurements on
colloidal systems~\cite{weeks00,kegel00} have shown that the dynamics is
heterogeneous. It was found, that the fastest particles within the system
are not distributed randomly but tend to form clusters. The size of these
clusters grows on lowering the temperature although by a factor that is
much smaller than the increase found in typical structural relaxation
times~\cite{donati99a}. Furthermore the precise connection between
heterogenous and cooperative dynamics is a question that is still open.

Since in most experimental setups it is not possible to access directly
the dynamics of the individual particles, people have proposed and
used methods to study the cooperative dynamics in supercooled liquids
in an indirect way. The idea is to investigate {\it confined} systems
and thus to introduce a new length scale, the thickness of the film
or the diameter of the pore, and to see how this scale influences
the relaxation dynamics. If CRR's do exist and grow with decreasing
temperature, the dynamics should differ from the bulk behavior as soon
as the size of the CRR's at a given temperature becomes comparable to
the system size~\cite{mrsconfit93,mrsconfit00,grenoble00,forrest01,grenoble03}.
One large class of experiments are done on molecular liquids (such as
salol, glycol, ortho-terphenyl) confined to porous material such as
Vycor glass, sol-gel glass and zeolithes with pore sizes ranging from
several hundred to only a few nanometer. Typical experimental methods are
differential scanning calorimetry (DSC)~\cite{jackson91,arndt97,hempel00},
dielectric spectroscopy~\cite{schueller94,pissis94,arndt96,kremer99,bergman00,fukao00},
neutron scattering~\cite{melnichenko95,zorn02}, solvation
dynamics~\cite{richert96,yang02,richert03} and many more.

As expected, most experimental data show a deviation from bulk
behavior below a certain threshold temperature. The main problem
with these experiments is the correct interpretation of the results,
which is not at all universal~\cite{mckenna00}.  Depending on the
investigated liquid, the realization of the confinement and even
the experimental method, there is a great variety in experimental
findings. Some give evidence for an accelerated dynamics in 
confinement~\cite{jackson91,pissis94,arndt96,fukao00,streck95,melnichenko95},
i.e. the glass transition temperature $T_g$ goes down, while others
see an increase of $T_g$~\cite{wallace95,schueller94,zheng95,huwe97}. 

In many cases in which the dynamics is slowed down, experimental
spectra suggest the existence of a secondary relaxation process
and often the interpretation of this result is based on the interaction between
liquid and surface. The idea is that there is an immobile
particle layer which is bound to the surface (slow process) and
the remaining liquid is then often accelerated with respect to the
bulk~\cite{wallace95,schueller94,streck95,melnichenko95,arndt96}.
A systematic study of supported and free standing polymer films
by ellipsometry~\cite{keddie94,forrest97,forrest00,dalnoki01} supports
this interpretation. While free standing films show a systematic decrease
of $T_g$ with decreasing thickness, the dynamics is partially slowed
down in the presence of an attractive surface (e.g. in supported
films). The experimental data can be described by a simple three
layer model~\cite{forrest00}, where one distinguishes between surface
layers for supported and free interfaces and a liquid like layer at
larger distance from the interface.  Other general trends found in
experiments include a broadening of the $\alpha-$relaxation peaks in
dielectric~\cite{pissis94,schueller95,melnichenko95,arndt97,morineau02}
but also in neutron scattering~\cite{zorn02} experiments.

As mentioned above, an important input parameter, which however is hard
to control in many experiments, is the interaction between surface and
liquid and the local structure. If the liquid is bound to the surface the
mobility is of course reduced and due to the existence of cooperativity
this slowing down could also propagate further into the system. The
opposite effect is seen, if the particle mobility is enhanced close
to the interface. Furthermore it is also important to know the precise
geometry of the confining material. E.g. in Vycor glass one could think
of a distribution of different pore sizes which leads to a superposition
of relaxation processes on slightly different time scales which could
explain the broadening of the relaxation peak for the whole system.

Since in an experiment only the {\it average} particle density can
be controlled, it is always possible that the {\it local} structure
shows strong deviations from bulk behavior, such as, e.g., density
oscillations. Lower local densities should of course accelerate
the dynamics and vice versa. That such oscillation in the density
do indeed occur is a result of many computer simulations of confined
liquids~\cite{bitsanis93,fehr95,nemeth99,delhommelle01,yamamoto00,varnik01,kranbuehl02,varnik02,binder03}.
Depending on the investigated liquid and the surface
interaction, the amplitude of such oscillations
can be very large~\cite{fehr95,nemeth99,yamamoto00},
but there are also situations where the density is almost
constant~\cite{boddeker99,teboul02,gallo03,kim03}. Thus it is not very
clear so far whether the change of the relaxation dynamics of confined
systems is a real effect of the confinement, or whether it is just a
``secondary'' effect, i.e. due to the interaction of the liquid with
the surface and hence the resulting change in structure and thus in
the dynamics.

Since in principle computer simulations offer the possibility to
control the most important input parameters such as the geometry of the
confinement and the interactions with the surface, and furthermore allow
to investigate the relaxation dynamics also on a local scale, they are
an ideal method to study the dynamics of confined systems and in the
present paper we report the results of such a study. In our simulation
we realize a situation where the static properties are {\it identical}
to the one in the bulk and therefore the dynamics is not influenced by
``secondary'' effects coming from a different structure. Thus this
helps to obtain a better understanding of the relaxation dynamics of
such confined systems and also to interpret some experimental findings.

In the following we first introduce a model for the investigated liquid and
the implementation of a rough wall with an amorphous structure.
Subsequently we investigate the dynamic properties using as observables
the mean squared displacement and the intermediate scattering function.
We characterize the influence of the wall on the local particle
dynamics and extract length scales of cooperativity from the structural
relaxation times. Finally we discuss finite size effects which show up
only at low temperatures or for very thin films, and analyze the average
relaxation dynamics, i.e. the only quantities that are accessible in a
real experiment.

\section{Model and details of the simulation}
\label{sec2}

We investigate a binary mixture of 80\% A and 20\%
B particles that interact via a Lennard-Jones (LJ)
potential $V_{\alpha\beta}(r)=4\epsilon_{\alpha\beta}
[(\sigma_{\alpha\beta}/r)^{12}-(\sigma_{\alpha\beta} /r)^{6}]$
($\alpha,\beta \in\{{\rm A,B\}}$).  The interaction parameters
are chosen as $\epsilon_{\rm AA}=1.0$, $\sigma_{\rm AA}=1.0$,
$\epsilon_{\rm AB}=1.5$, $\sigma_{\rm AB}=0.8$, $\epsilon_{\rm BB}=0.5$, and
$\sigma_{\rm BB}=0.88$. The potential is truncated and shifted at cut-off
radii $r_{\alpha\beta}^c$$=2.5\cdot \sigma_{\alpha\beta}$.  Within this
paper all results will be given in reduced LJ  units, i.e. length
in units of $\sigma_{\rm AA}$, energy in units of $\epsilon_{\rm AA}$, and
time in units of $(m\sigma_{\rm AA}^2/48\epsilon_{\rm AA})^{1/2}$. The
introduction of two different length scales for the size of the
particles strongly suppresses the crystallization of the liquid
at low temperatures. Previous simulations in the bulk have shown
that this system is a very good glass-former, i.e. is not prone to
crystallisation~\cite{kob95a,kob95b}. As a reference for the relevant
temperature scale we note that the system is in a normal liquid
state for $T=1.0$, that the critical temperature of the mode-coupling
theory~\cite{mct} is around 0.435, and that the Kauzmann temperature
is around 0.3~\cite{kob95a,kob95b,coluzzi00,mezard00,sciortino01}

The equations of motion have been integrated with the velocity form of
the Verlet algorithm, using a step size of 0.01 and 0.02 for $T\geq 1.0$
and $T<1.0$, respectively. During these runs the accessible volume to
the particles was keep constant.

As already mentioned in the introduction, it is most important to be
able to decide whether the change in the relaxation dynamics is due to
the {\it confinement} or whether it is related to the {\it interaction}
of the fluid particles with the wall, i.e. a change of the local
structure close to the wall. Since in the present work we want to study
only the influence of the confinement on the dynamics, we have to
avoid a change of the structural properties due to the confining walls.

In Ref.~\cite{scheidler00} we presented a simulation of a liquid
confined in a tube in which the positions of the wall particles were
given by a frozen configuration of the same liquid at a {\it fixed}
temperature $T_{\rm W}$ (which corresponded to a temperature at which
the bulk liquid was slightly supercooled). Since the structure of the
wall was thus almost identical to the enclosed liquid in its bulk phase,
the difference of the structure of the confined liquid from the one of
the bulk was fairly small. Nevertheless, due to the slightly different
structure, the confined liquid was not quite in equilibrium after the
introduction of the walls and hence it was necessary to equilibrate the
confined system.  Due to the strong slowing down of the confined liquid
close to the wall, this equilibration took a very long time and hence it
was possible to access only intermediate temperatures, i.e. the relaxation
dynamics of the confined system at low temperatures could not be studied.

It is, however, possible to avoid this problem by setting the
``wall temperature'' $T_{\rm W}$, i.e. the temperature at which
this configuration was equilibrated, identical to the temperature of
the contained liquid~\cite{parisi}. In contrast to the simulation of
Ref.~\cite{scheidler00} the wall structure is now also a (weak) function
of temperature. Furthermore it is necessary to add a hard core potential
directly at the interface in order to prevent that the fluid particles
penetrate into the wall. In the following we will briefly demonstrate that
with this setup the structure of the confined liquid is indeed identical
to the one of the bulk system. Although we will restrict ourself to the
situation in which the liquid is confined between two parallel plane
walls, the derivation can easily be expanded to an arbitrary confining
geometry.  Due to this fact it is hence not necessary to equilibrate
the liquid and thus the above mentioned problem is avoided.

Consider a static variable $X({\bf R})$ where ${\bf R}$ is a
$3N$-dimensional vector that characterizes the position of all particles
in a bulk system. In the canonical ensemble the expectation value of
the observable $X$ is given by:

\begin{equation}
\langle X \rangle = 
    \frac{1}{Z_N} \int {\rm d}^{3N}{\bf R}  \, X({\bf R}) \,
\exp \left[ -\beta U({\bf R}) \right]  \, ,
\label{eq1}
\end{equation}

\noindent
where $U({\bf R})$ is the total potential between all particles and $Z_N=
\int {\rm d}^{3N}{\bf R} \, \exp [-\beta U({\bf R})]$ the partition
function. The configuration average in Eq.~(\ref{eq1}) can be split
into contributions that come from domains in which the particles are
inside the fluid region, denoted by $F$, and outside of it (denoted by
$W$). (Note that at this stage this division is purely formal.):

\begin{equation}
\langle X \rangle = 
    \frac{1}{Z_N} 
\left[ \int_F {\rm d}{\bf r}_1 +\int_W {\rm d}{\bf r}_1\right] \ldots
\left[ \int_F {\rm d}{\bf r}_N +\int_W {\rm d}{\bf r}_N\right]
  \, X({\bf R}) \, \exp \left[ -\beta U({\bf R}) \right] .
\label{eq2}
\end{equation}

If we now relabel the particles such that the ones that are in the
wall-domain have index $1\ldots k$, with $k \in \{0,...,N\}$, and thus the
remaining $N-k$ particles are in the fluid-domain, we can write $\langle
X \rangle$ as follows:

\begin{equation}
\langle X \rangle =           
    \frac{1}{Z_N} \sum_{k=0}^N
    \int_W {\rm d}{\bf r}_1 \ldots \int_W {\rm d}{\bf r}_k
    \int_F {\rm d}{\bf r}_{k+1} \ldots \int_F {\rm d}{\bf r}_N
    \, X({\bf R}) \, \exp \left[ -\beta U({\bf R}) \right] .
\label{eq3} 
\end{equation}

In the following we will denote by $\alpha (k)$ the $3k$ dimensional
vector that describes a configuration of $k$ particles that characterize
a given wall. The canonical average of an observable $X$ that depends
only on the liquid particles in the fluid domain and that is confined
by a wall given by the $k$ particles at position $\alpha (k)$ is given by:

\begin{equation}
\langle X \rangle_{\alpha (k)} = \frac{1}{Z_\alpha}
     \int_F {\rm d}{\bf r}_{k+1} \ldots {\rm d}{\bf r}_N
      X({\bf r}_{k+1} \ldots {\rm d}{\bf r}_N) \, 
      \exp\left[ - \beta U_\alpha({\bf r}_{k+1} \ldots {\bf r}_N) \right] .
\label{eq4}
\end{equation}

\noindent
Here $U_\alpha({\bf r}_{k+1} \ldots {\rm d}{\bf r}_N)$ is the potential
energy of the system if the position of the first $k$ particles are given
by $\alpha (k)$, and $Z_\alpha$ is the partition function for the liquid
system with walls $\alpha (k)$. (Note that here we have made use of the
fact that we have at the interface a hard core potential and that thus
the domain of integration is restricted to $F$.)

To determine the expectation value of an observable $X$ for a confined
system, one has to average over all realizations of the walls:

\begin{equation} 
\langle X \rangle_F = \sum_{k=0}^{N} \sum_{\alpha(k)} P(\alpha(k)) \langle X \rangle_{\alpha (k)}
\label{eq5}
\end{equation}

\noindent
where $\sum_{\alpha(k)}$ stands for the sum over all possible wall
configurations with $k$ particles and $P(\alpha(k))$ is the statistical
weight of such a configuration. This weight is just given by the trace
of the Boltzmann factor over the fluid particles:

\begin{equation}
P(\alpha(k)) = \frac{1}{Z_N} \int_F {\rm d}{\bf r}_{k+1} \ldots \int_F {\rm d}{\bf r}_N
    \, \exp \left[ -\beta U_\alpha({\bf r}_{k+1} \ldots {\bf r}_N) \right] .
\label{eq6}
\end{equation}

By means of Eqs.~(\ref{eq3})-(\ref{eq6}) it is now easy to see that we
have $\langle X \rangle_F= \langle X \rangle$, i.e. that by averaging
over all possible walls the static properties of the confined system
are the same as the one of the bulk.

In practice we proceeded as follows to determine the dynamical properties
of the confined liquid in a film geometry with thickness $D$: First we
equilibrated a rectangular ($L\times L \times D$) bulk configuration
with $L=12.88$.  To create the film of thickness $D$ (in $z-$direction,
$0 < z < D$) we introduced two rigid walls, at $z=0$ and $z=D$. The
configuration for the wall particles is given by the periodic images
of the same configuration which are translated in $z-$direction by $D$
and $-D$.  Since we use cut-off radii, only wall particles within a
slice of thickness $r_{\rm AA}^{c}=2.5$ have to be taken into account.
The liquid interaction and also the average particle density of
$\rho=1.2$ is identical to the bulk simulations that have been done
earlier~\cite{kob95a,kob95b}. Most of the simulations have been done
for $D=15.0$, although thinner films have been considered as well
(see below). (Note that for $D=15.0$ the given density corresponds to
3000 particles.) We have found that at low temperatures the relaxation
dynamics shows strong finite size effects, see Section~\ref{sec3f}.
Therefore we restricted our investigations to the temperature range $2.0
\geq T \geq 0.5$ for which such effects are not present.

The remarkable advantage of the just discussed method to confine
the liquid is that we only have to equilibrate bulk systems where
relaxation times are orders of magnitude smaller than in films with
rough surfaces~\cite{scheidler00}. Of course the production run has
to be much longer than the equilibration runs in order to investigate
the structural relaxation also close to the surface. A remaining
problem is the realization of the average over the different walls,
see Eq.~(\ref{eq5}). In practice we calculated this mean by averaging
over 16 independent walls, a number that, for the system sizes used in
the present simulation, turned out to be sufficient.  In particular
we found that the average particle density as a function of $z$ is a
constant, as it has to be for a homogeneous system and that structural
quantities like the partial structure factors are identical to the ones
in the bulk~\cite{scheidler_diss}.

\section{Relaxation dynamics}
\label{sec3}

Since we already know from previous simulations that the dynamic
properties will strongly depend on the particle distance $z$
from the wall~\cite{scheidler00,scheidler03} we will concentrate
on observables that take into account this spatial dependence. In
particular we will investigate the single particle dynamics such as mean
squared displacements and the self part of the van Hove correlation
functions. Subsequently we will use the decay of density fluctuations
to extract characteristic relaxation times and study their temperature
dependence. Finally we will study collective quantities and dynamic
properties averaged over all particles.

\subsection{Local single-particle dynamics}
\label{sec3a}

One very useful observable to characterize the dynamics
of a tagged particle is the mean squared displacement
(MSD)~\cite{hansen86,barrat03}. Since this function is expected to depend
on $z$ we generalize the usual definition of the MSD in the following way:

\begin{equation}
r^2_\alpha (z,t) = \frac{1}{N_{\alpha} } \sum_{k=1}^{N_{\alpha}}
           \langle |{\bf r}_k(t) -{\bf r}_k(0)|^2
           \delta ( z_k(0)-z )  \rangle \, .
\label{eq7}
\end{equation}

\noindent
Thus $r^2_\alpha (z,t)$ corresponds to the dynamics of particles of
type $\alpha \in \{\rm A,B\}$ which at time $t=0$ had a distance $z$
to one of the walls. (Note that for $t>0$ the particle will change its
distance from the wall, i.e. $z(t) \neq z(0)$ for $t>0$.) Furthermore
we will see that it is useful to distinguish between displacements
in different directions, namely $r^{2,z}_\alpha (z,t)$ perpendicular
to the wall and $r^{2,p}_\alpha (z,t) = r^{2,x}_\alpha (z,t) +
r^{2,y}_\alpha (z,t)$ for displacements parallel to the film plane.
To calculate $r^2_\alpha (z,t)$ we divided the film into 10 slices
(of thickness $D/10=1.5$) and averaged over the two slices that have
the same distance to one of the walls.

\begin{figure}
\includegraphics[width=70mm]{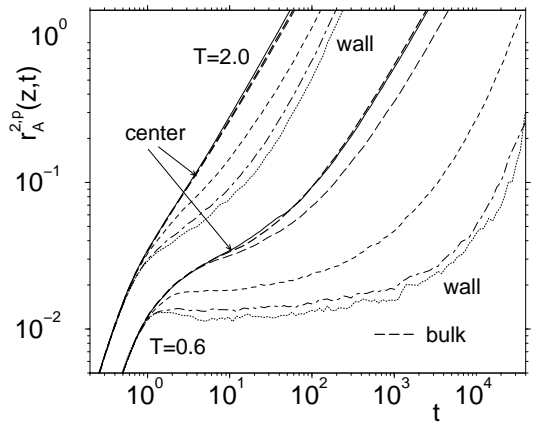}
\caption{\label{fig_msd_p_center_vs_wall} 
Parallel component of the mean squared displacement $r^{2,p}_\alpha (z,t)$
for A particles in layers with different distance to one of the walls
($z=0.75, 2.25, 3.75, 5.25, 6.75$) at $T=2.0$ and $T=0.6$.  The bold
dashed lines are the bulk data at the same temperatures.}
\end{figure}

In Fig.~\ref{fig_msd_p_center_vs_wall} we show the time dependence of
$r^{2,p}_\alpha (z,t)$ for two temperatures and the five values
of $z$ (see caption for details). Also included in the figure are the
corresponding curves for the bulk (bold dashed lines) and in order
to understand the data for the confined system it is useful to start
the discussion with their time dependence. At short times the bulk
curves show a $t^2$ dependence since on that time scale the particles
move just ballistically. At high $T$ this time dependence crosses over
immediately to the diffusive regime, i.e. the MSD increases linearly with
time. The same short and long time behavior is found also at low $T$.
However, we see that at these temperatures the two regimes are separated
by a time-window in which the MSD increases only slowly. The physical
reason for this behavior is the so-called ``cage-effect'', i.e. the
fact that at intermediate times the particles are temporarily trapped
by their surrounding neighbors. This time window is often also called
``$\beta-$relaxation''~\cite{mct}.

The curves for the confined system for particles that are in the
center of the film (thin lines) are basically identical to the one
of the bulk system, showing that for large distances from the wall
the relaxation dynamics is not changed. This is in contrast to the
dynamics of the particles close to the walls, dotted lines, which seems
to be significantly slower than the one of the bulk.  Already at high
temperatures we see that the MSD has a weak plateau at intermediate times,
thus demonstrating that the cage-effect is already present. This becomes
much more pronounced if the temperature is lowered. E.g. we find that
at $T=0.6$ the MSD for the smallest $z$ shows a plateau that extends
over several decades in time and only at the very end of the run the
particles start to show a diffusive behavior.  (Note that at very long
times the MSD for the different values of $z$ must come together since
ultimately the system will be ergodic and hence the dynamical properties
of a particle cannot depend on its location at time zero. However, for
the lowest temperatures considered these ``mixing'' times are well beyond
the time scale of our simulation.) Thus we conclude from the figure that
the mobility of a particle at the surface is strongly suppressed and
that the typical time scale to leave its cage is orders of magnitude
larger than the one of a particle in the bulk.

\begin{figure}
\includegraphics[width=73mm]{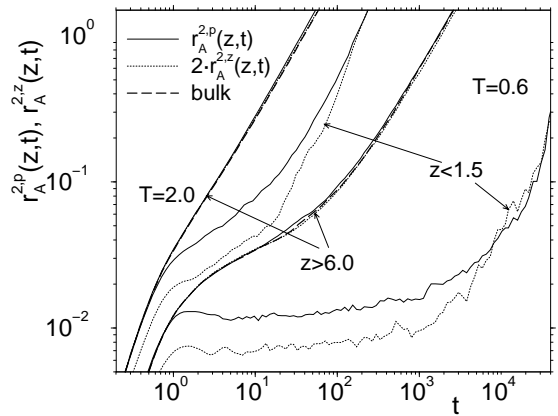}
\caption{\label{fig_msd_isotropy}
Anisotropy of the mean squared displacement for A particles at $T=2.0$
and $0.6$.  Displacements $r^{2,p}_{\rm A} (z,t)$ in film plane and
$r^{2,z}_{\rm A} (z,t)$ perpendicular for particles with $z \ge 6.0$
(center of film, curves close to bulk data) and $z \le 1.5$ (region
close to the surface). Also shown are the isotropic bulk data (bold
dashed lines).}

\end{figure}

The curves shown in Fig.~\ref{fig_msd_p_center_vs_wall} are the component
of the MSD that is parallel to the confining wall. It can be expected,
however, that the relaxation dynamics is not isotropic and that therefore
it is of interest to investigate also the component of the motion
perpendicular to the wall. This is done in Fig.~\ref{fig_msd_isotropy}
where we show for two temperatures and two values of $z$ the components
$r^{2,p}_\alpha (z,t)$ and $r^{2,z}_\alpha (z,t)$. From this graph
we see that for $z\geq 6.0$, i.e. in the center of the film, the parallel
and perpendicular component are the same and that they are identical
to the one in the bulk (bold dashed lines). For $z\leq 1.5$ this is,
however, not the case in that both components show a significantly
slower $t-$dependence than the one of the bulk. Furthermore we see that
the relaxation in $z$ direction is at intermediate time scales slower
than the one in $x-y$ direction. Also the size of the cage in $z$
direction is significantly smaller than the one in $x-y$ direction,
as can be recognized by the fact that the height of the plateau at
intermediate times is smaller for $r^{2,z}_\alpha (z,t)$ than the one
seen in $r^{2,p}_\alpha (z,t)$. This observation is related to the
fact that the particles of the wall are completely frozen, i.e. do not
oscillate around an ``equilibrium'' position, and hence are not able to
make space for the vibrations of the fluid particles.

In order to investigate the relaxation dynamics of the particles in more
detail it is useful to investigates not only their averaged displacements
but their distribution at different times, i.e. to study the self part
of the van Hove correlation function~\cite{hansen86,barrat03}. Also here
we have to generalize this function in order to take into account its
dependence on $z$:

\begin{equation}
G_{\rm s}(z,{\bf r},t) = \frac{1}{N_{\alpha}} \sum_{k=1}^{N_{\alpha}}
\langle \delta ( {\bf r} - [{\bf r}_k(t) - {\bf r}_k(0)] )
\delta(z_k(0)-z) \rangle \, .
\label{eq8}
\end{equation}

In order to improve the statistics of our data, we do not distinguish
between the different orientations of the vector $\bf r$ and instead
average over the angular dependence of $G_{\rm s}(z,{\bf r},t)$. (See
Ref.~\cite{scheidler_diss} for a discussion of this angular dependence.)

\begin{figure}
\includegraphics[origin= 0 140, width=73mm]{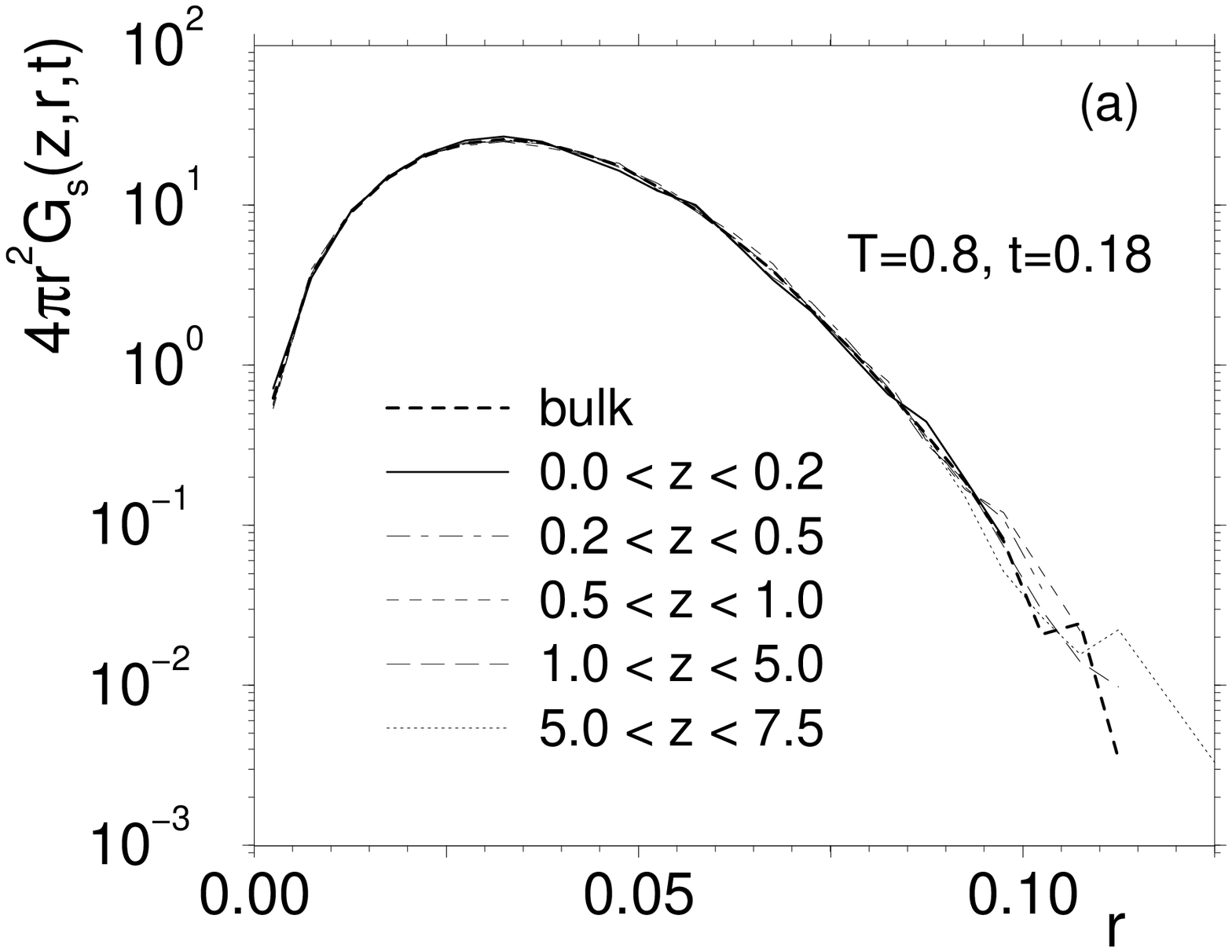}
\includegraphics[origin= 0 70, width=73mm]{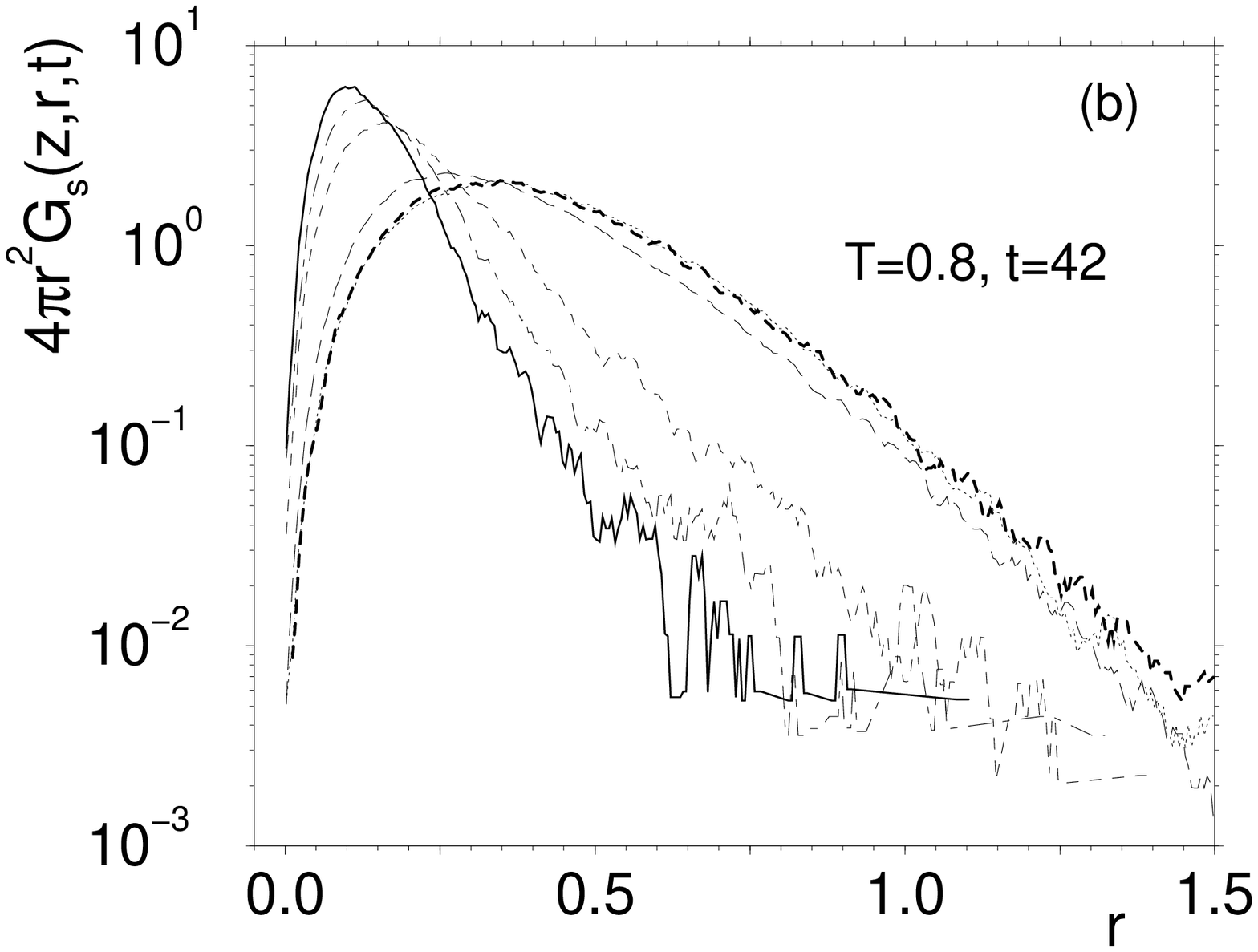}
\includegraphics[origin= 0 0, width=73mm]{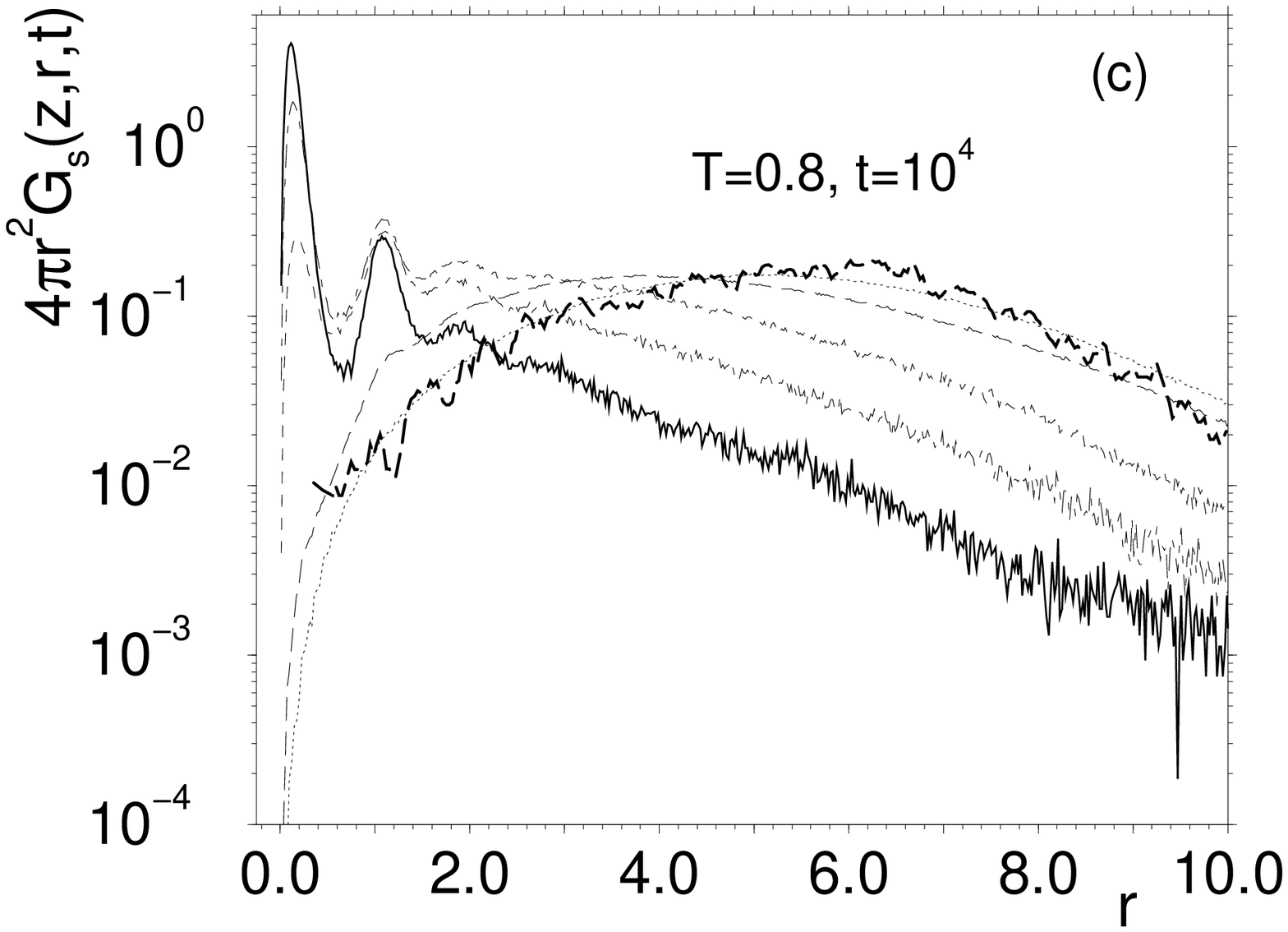}
\caption{\label{fig_vanhove_layerdependence}
Self part of the generalized van Hove correlation function for the A particles (see
Eq.~(\ref{eq8})) for different values of $z$. The bold dashed line is the
distribution for the bulk system. The different line styles correspond
to different values of $z$ (see panel (a)). $T=0.8$ a) $t=0.18$; b)
$t=41$; c) $t=10^4$.}
\end{figure}

The $r-$dependence of $4\pi r^2 G_{\rm s}(z,r,t)$ for the A particles is shown in
Fig.~\ref{fig_vanhove_layerdependence} for three characteristic times.
At the first time, $t=0.18$, the particles are still in the ballistic
regime and thus the distribution is independent of $z$, see
Fig.~\ref{fig_vanhove_layerdependence}a.

Figure \ref{fig_vanhove_layerdependence}b shows that for a time in
which the bulk system is in the $\beta-$relaxation regime, the van Hove
function does depend strongly on the distance from the wall. For large
values of $z$, and hence also for the bulk system, the distribution has
a peak around 0.3 and non-negligible values even for $r$ larger than 1.0,
i.e. the distance that corresponds to the nearest neighbor distance. This
means that these particles have left the cage in which they were at
time $t=0$. In contrast to this, the curves for small values of $z$
are still localized at distances around 0.1, i.e. the particles close
to the wall have not yet left their initial cage.

Only at very long times a substantial number of particles
at the surface start to leave their original cage, see
Fig.~\ref{fig_vanhove_layerdependence}c. From the presence and the nature
of the time evolution of the secondary peak in the distribution function
(located around $r=1.0$) we conclude that the responsible mechanism is a
jump type of motion, i.e. these particles do not move continuously but hop
to a new position that has a typical distance 1.0 from the initial site.
In the bulk system such a hopping motion has been observed only for the
B particles and even this only at very low temperature~\cite{kob95a}.
In general such type of relaxation dynamics is more characteristic for
strong glass formers, such as, e.g., silica~\cite{horbach99}. In these
systems the typical ``jump distance'' corresponds to the first peak
in the radial distribution function and it is possible to recognize
even the existence of peaks of higher order and which correspond to
subsequent jumps.

The reason why the confined system shows a hopping motion that is
not present in the bulk system is related to the presence of the
walls. Since these walls consist of immobile particles they give
rise to a (static!) local potential energy landscape with energy
minima for positions in between those particles and the observed
jumps are just the motion of the particles between these local
minima. Note that these jumps are not necessarily parallel to
the surface and are not confined to the first layer of the fluid
since they are also observed for particles that are in the second
and third layer (see Fig.~\ref{fig_vanhove_layerdependence}c and
Ref.~\cite{scheidler_diss}). The reason for this is that the above
mentioned energy landscape propagates also somewhat into the liquid and
hence affects the relaxation dynamics also at larger values of $z$.

We now address the question how to quantify the slowing down of the
dynamics close to the surface. As we will see in the following it is
possible to extract from the decay of time dependent correlation functions
characteristic relaxation times which show a pronounced $z$-dependence.
We will concentrate on wave vector dependent intermediate scattering
functions in reciprocal space~\cite{hansen86}, but the investigation of
similar density correlators in real space shows the same qualitative
and even quantitative results~\cite{scheidler_diss}. Here we will consider the
two following functions:

\begin{equation}
F({\bf q},z,t)= \frac{1}{N_\alpha}
       \sum_{j=1}^{N_\alpha}  \sum_{k=1}^{N_\alpha}
          \langle \exp [ i {\bf q} \cdot ( {\bf r}_k(t) - {\bf r}_j(0) ) ]
          \delta(z_j(0) - z) \rangle \, ,
\label{eq10}
\end{equation}

\noindent
and its diagonal part

\begin{equation}
F_s({\bf q},z,t)= \frac{1}{N_\alpha}
       \sum_{j=1}^{N_\alpha} 
          \langle \exp [ i {\bf q} \cdot ( {\bf r}_j(t) - {\bf r}_j(0) ) ]
          \delta(z_j(0) - z) \rangle \, .
\label{eq11}
\end{equation}

\noindent
Thus $F({\bf q},z,t)$ and $F_s({\bf q},z,t)$ are just the coherent and
incoherent intermediate scattering functions generalized in such a way
to take into account their $z-$dependence. Note that, in contrast to the
case of a bulk liquid that is isotropic, we have to take into account
the orientational dependence of these functions. In the following we
will concentrate on $F^p(q,z,t)$ and $F_s^p(q,z,t)$, the functions in
which one has taken the angular average over $\bf q$ for wave-vectors
that are parallel to the walls. Although most of the results presented
below are for the A particles and the wave-vector $q=7.2$, the location
of the maximum in the partial A-A structure factor~\cite{kob95b}, we
have found qualitatively similar results for the B particles and/or for
other wave-vectors~\cite{scheidler_diss}.

\begin{figure}
\includegraphics[width=85mm]{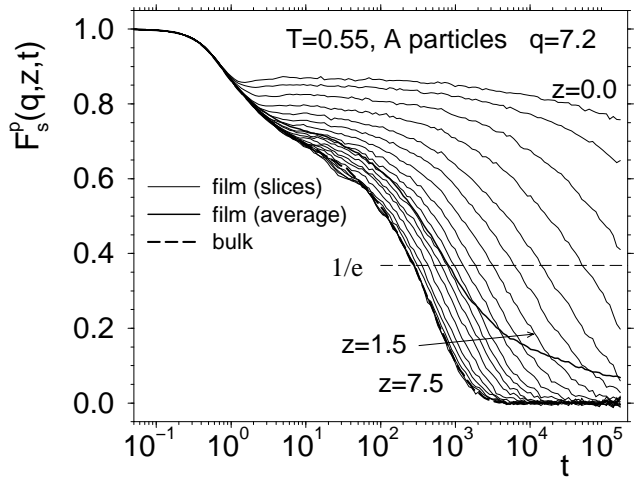}
\caption{\label{fig_fsqzt}
Self part $F^p_{\rm s}(q,z,t)$ of the intermediate scattering
function at $T=0.55$ and $q=7.2$ for A particles. The different
curves correspond to the following distances from the wall: $z=0.0, 0.25,
0.5, \ldots, 2.75, 3.0, 3.5, 4.0, 4.5, 5.5, 6.5, 7.5$). The bold solid curve
is the average over the whole film and the bold dashed curve is the
bulk curve. The horizontal dashed line is used to define a relaxation
time $\tau$.
}
\end{figure}

Let us first consider the typical behavior of the incoherent intermediate
scattering function at low temperatures in bulk systems (dashed
curve in Fig.~\ref{fig_fsqzt}). The decay of $F^p_{\rm s}(q,t)$
happens in two steps, first the decay to a plateau which corresponds
to the relaxation inside the cage, see the above discussion of the
mean squared displacement, and a second step that corresponds to the
$\alpha-$relaxation. Note that for $T=0.55$ the mentioned plateau is
not yet very well developed, but its width increases rapidly with
decreasing temperature~\cite{kob95b}.

As expected, $F^p_{\rm s}(q,z,t)$ for particles in the center of the film
are identical to the bulk curve. Upon approach to the wall we recognize
again the continuous slowing down of the dynamics. Compared to
the $z-$dependence of the slowing down found in the mean squared
displacements the effect is even more pronounced, since here the time
scales change by many decades. (Note that for the smaller values of
$z$ considered, the correlators do not decay to zero anymore since
the dynamics is so slow. Hence we see that it is indeed crucial to
have starting configurations that are already equilibrated in order to
avoid aging effects.) From the figure we also recognize that the reason
for the unusual shape of the {\it average} correlator, bold solid curve in
Fig.~\ref{fig_fsqzt}, is that the curves close to the wall are decaying
so much slower than the one in the center of the film.

\begin{figure}
\includegraphics[width=85mm]{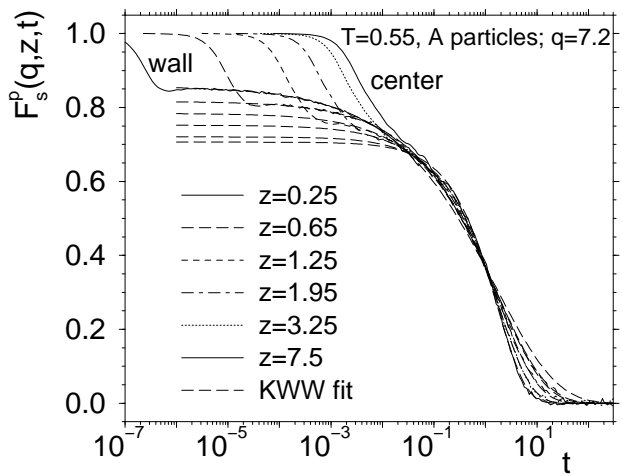}
\caption{\label{fig_fsqzt_superpos}
Time dependence of $F^p_{\rm s}(q,z,t)$ as a function of rescaled
time $t/\tau$ (in bold).  The values of $z$ are $z=0.25, 0.65, 1.25, 1.95,
3.25$, and 7.5. The thin dashed lines are fits to the data in the $\alpha-$relaxation
regime with a KWW-law (see Eq.~(\ref{eq12})).
}
\end{figure}

Figure~\ref{fig_fsqzt} shows that the typical time scale for the
relaxation increases rapidly for decreasing $z$. The question arises
whether it is only the time scale that depends on $z$ or whether also
the shape of the curve depends on the distance from the wall. To check
this we can define a relaxation time $\tau$ by requiring that at this
time the correlator has decayed to $1/e$ (see horizontal dashed line in
Fig.~\ref{fig_fsqzt}). In Fig.~\ref{fig_fsqzt_superpos} we show a plot
of $F^p_{\rm s}(q,z,t)$ vs. $t/\tau$ for various values of $z$. If
the shape of the curves would be independent of $z$ we would find that
this rescaling of time produces a master curve. However, from the figure we see
that this is not the case since the correlators for small values of $z$
are more stretched.

In order to quantify this effect we have fitted the correlators by a
Kohlrausch-Williams-Watts function (KWW),

\begin{equation}
F^p_{\rm s}(q,z,t)=f_c(z) \cdot
\exp \left[  -\left( \frac{t}{\tilde{\tau}(z)} \right)^{\beta(z)} \right] \, ,
\label{eq12}
\end{equation}

\noindent
a functional form often used to describe the $\alpha-$relaxation
of supercooled liquids. That this function is able to characterize
well also the correlators in the confined liquid is demonstrated in
Fig.~\ref{fig_fsqzt_superpos} where we have included the KWW fits to
the data as well (thin dashed lines).

\begin{figure}
\includegraphics[width=85mm]{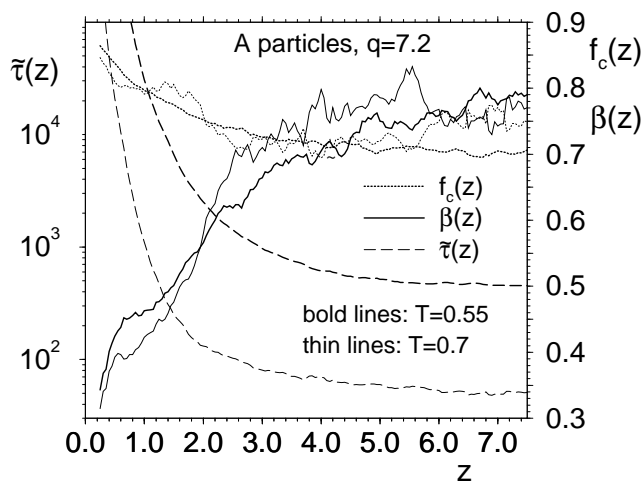}
\caption{\label{fig_fsqzt_kwwpar}
$z-$dependence of the KWW parameters $f_c$, $\tilde{\tau}$ and $\beta$
as obtained from fits to $F^p_{\rm s}(q,z,t)$ for $T=0.7$ (thin
lines) and $T=0.55$ (bold lines).
}
\end{figure}

In Fig.~\ref{fig_fsqzt_kwwpar} we show the $z-$dependence of the
fit parameters obtained by fitting $F^p_{\rm s}(q,z,t)$ with a KWW
function. The two set of curves correspond to $T=0.7$ and $T=0.55$
(thin and bold lines, respectively).  We start the discussion with the
relaxation times $\tilde{\tau}$ (dashed lines in the figure). From
the figure it becomes clear that for large $z$ this relaxation time
is basically independent of $z$ and is the same as the one found in
the bulk. For decreasing $z$ it starts to rise rapidly and at the
smallest $z$ for which we were able to determine it, it is more than
two decades higher than the bulk value. The parameter $f_c$, i.e. the
height of the plateau in the $\beta-$relaxation regime, shows only
a relatively mild $z-$dependence in that it increases by about 20\%
in the $z-$range considered (dotted curves). In contrast to this, the
third parameter of the KWW-function, the stretching parameter $\beta$,
shows a relatively strong $z-$dependence in that it decreases from
a value around 0.8 to $\beta \approx 0.33$ at small $z$. Thus we see
that not only do the relaxation times increase with decreasing $z$, but
that also the stretching increases significantly. The likely reason for
this strong stretching is that very close to the surface the system is
very heterogeneous in the sense that a slight modification of the local
(frozen) environment, and such variations are certainly present, will
give rise to very strong fluctuations in the relaxation behavior. Hence
if one averages over these fluctuations the result is a very pronounced
stretching. Note that this small value of $\beta$ is also the reason
why the correlators for small $z$ decay by 4-5 orders of magnitude
slower than the ones at large $z$ (see Fig.~\ref{fig_fsqzt}), whereas
the relaxation time $\tilde{\tau}$ increases only by a factor of $10^2$.

A comparison of the data sets for $T=0.7$ and $T=0.55$ shows that the
$z-$dependence for $f_c$ and $\beta$ are, within the noise of the data,
the same. This implies that also for the confined system the so-called
time-temperature superposition is valid, i.e. that the shape of the
curves is independent of temperature. This feature is often found in
glass forming liquids and in particular also for the present binary
Lennard-Jones system~\cite{kob95b}. However, we already point out at
this place that the relaxation times for the two temperatures do {\it
not} differ by just a constant factor. We will come back to this point
in Sec.~\ref{sec3c}.

\subsection{Factorization property of generalized correlation functions}
\label{sec3b}

Since the relaxation dynamics of the present Lennard-Jones system in
the {\it bulk} is described well by means of mode-coupling theory, we will try
to test to what extend some of the predictions of this theory hold for
our confined systems. We have already seen that for a given value of $z$
the time-temperature superposition principle holds, in agreement with the
prediction of MCT. A further important prediction of the theory is the
so-called factorization property for the $\beta-$relaxation: Consider
any correlator $\Phi_x(t)$, where $x$ stands for wave-vector, particle
species, etc. MCT predicts that in the time window of the $\beta-$regime
the time dependence of $\Phi_x(t)$ can be written as follows:

\begin{equation}
\Phi_x(t) = f^{\rm c}_x + h_x G(t)  \, .
\label{eq13}
\end{equation}

\noindent
Here $f^{\rm c}_x$ and $h_x$ are the height of the plateau and an
amplitude, respectively. The remarkable statement of MCT is that the
whole time and $T-$dependence of the correlator is given by the function
$G(t)$, the so-called $\beta-$correlator, and that this function is
system universal, i.e. it does not depend on $x$.

One possibility to check whether or not the factorization property holds
is to calculate the following function~\cite{signorini90}:

\begin{equation}
R_x(t)= \frac{\Phi_x(t) - \Phi_x(t')}{\Phi_x(t'') - \Phi_x(t')} \, ,
\label{eq14}
\end{equation}

\noindent
where $t'$ and $t''$ are two arbitrary times in the $\beta-$regime.
If the factorization property holds, the function $R_x(t)$ should be
independent of $x$.

For the bulk system it was found that the factorization property
does indeed hold very well in that the data for $R_x(t)$ for
all sort of different correlators do indeed fall on top of each
other~\cite{gleim99}. In the following we thus will check to what extend
this holds also true for $z-$dependent correlators.

\begin{figure}
\includegraphics[width=85mm]{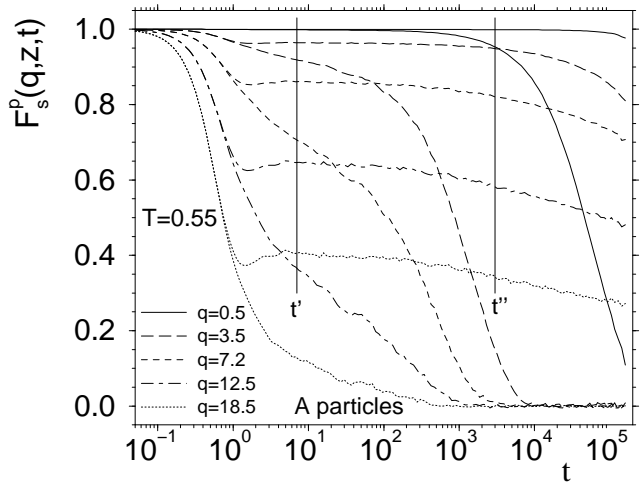}
\caption{\label{fig_fsqzt_qdep}
Time dependence of $F^p_{\rm s}(q,z,t)$ for the A particles at
$T=0.55$ for different wave-vectors (see legend) for particles in the
center ($7.35<z<7.5$, bold lines) and at the surface ($0 < z < 0.15$,
thin lines).
}
\end{figure}

To begin we investigate the wave-vector dependence of these functions.
Figure~\ref{fig_fsqzt_qdep} shows that qualitatively the time dependence
of the correlators is independent of $q$ but that the typical relaxation
time as well as the height of the plateau increases rapidly with decreasing
$q$. For large values of $z$ (bold curves) and large wave-vectors it is
in fact difficult to see that there is indeed a plateau (see curve for
$q=18.5$) since the curve decays very quickly. This is due to the fact
that large wave-vectors correspond to small displacements of the order
$2 \pi/q$ and that thus a decorrelation of the particle position can
take place already on a very short time scale. No such fast decay is
seen for small values of $z$ (thin curves) since there the relaxation
dynamics is very slow even on small length scales.

\begin{figure}
\includegraphics[origin= 0 185, width=80mm]{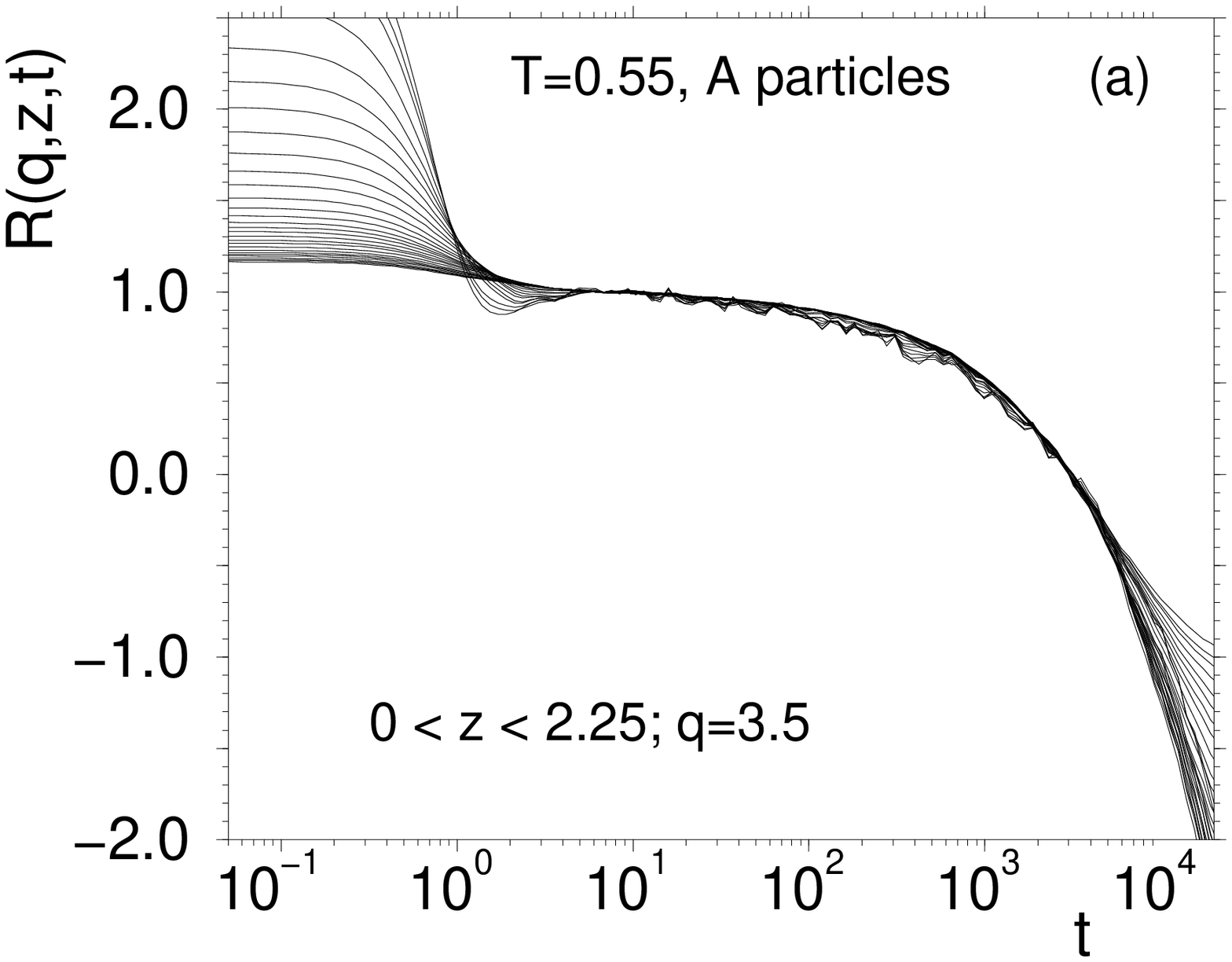}
\includegraphics[origin= 0 130, width=80mm]{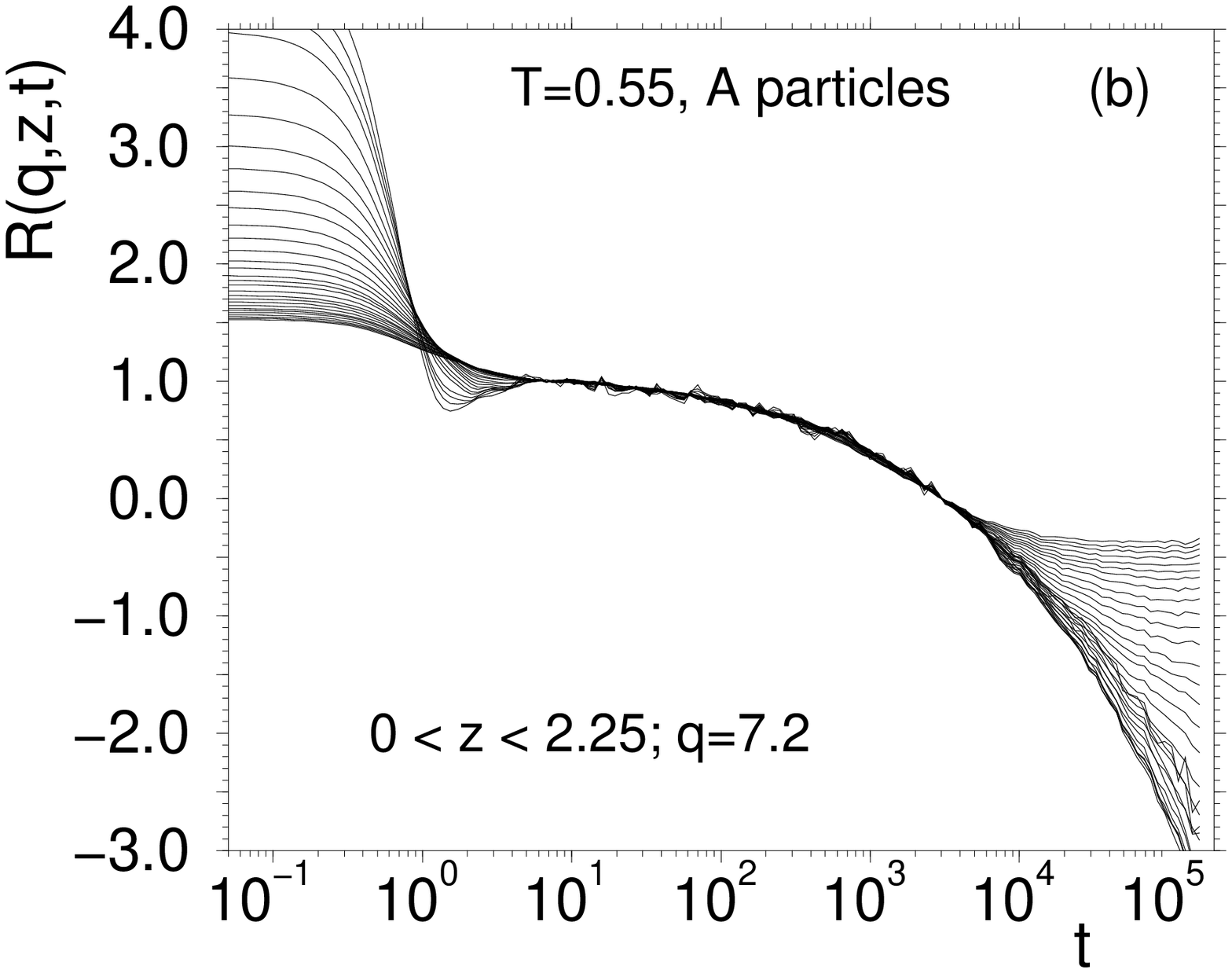}
\includegraphics[origin= 0 75, width=80mm]{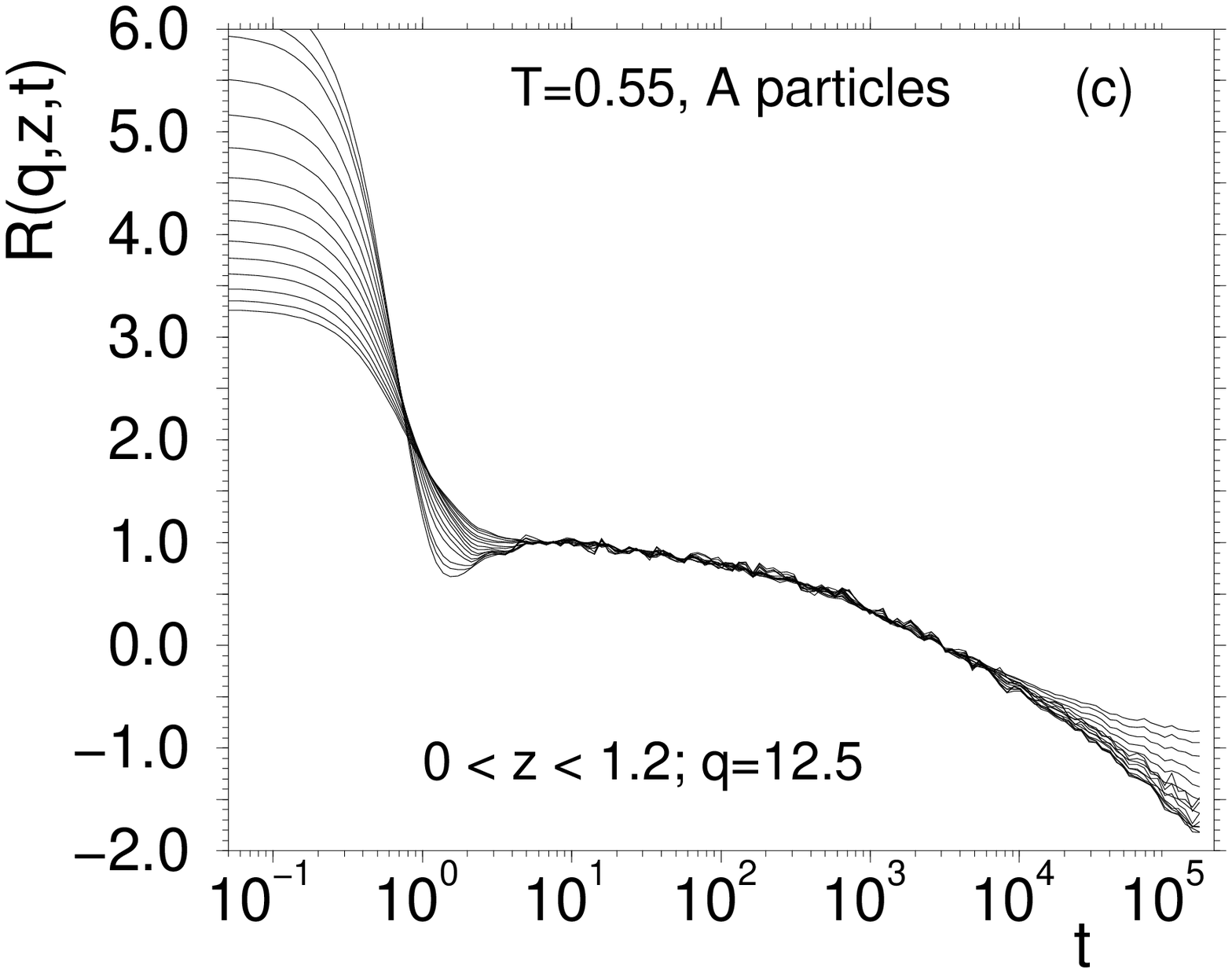}
\includegraphics[origin= 0 0, width=80mm]{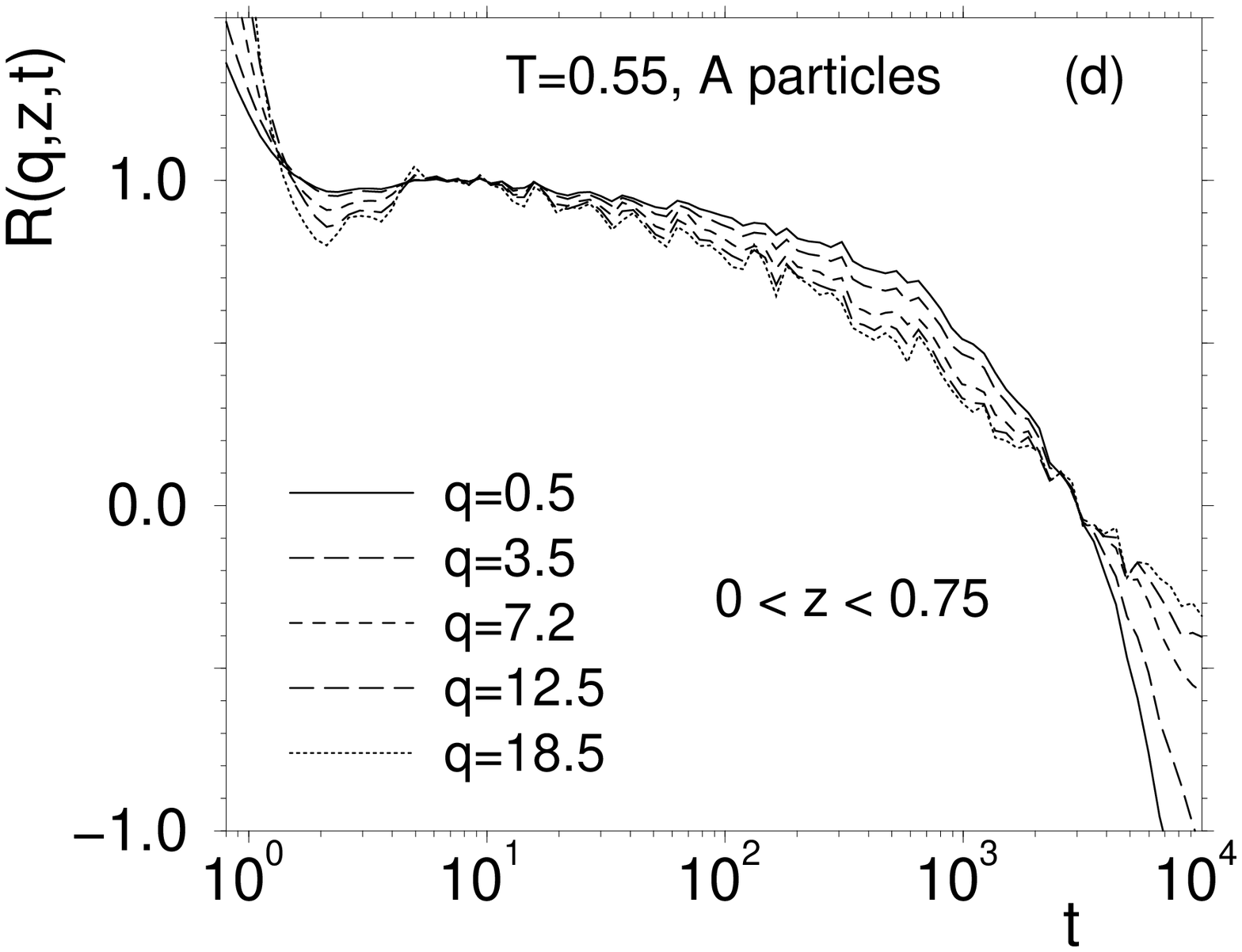}
\caption{\label{fig_fsqzt.factorization}
Time dependence of the function $R(q,z,t)$ from Eq.~(\ref{eq14}) in
order to test the factorization property for the incoherent intermediate
scattering function $F^p_{\rm s}(q,z,t)$ for A particles and different
values of $z$ (see main text for details). (a) $q=3.5$; (b) $q=7.2$;
(c) $q=12.5$. (d) The master functions $R(q,z,t)$ for different values
of $q$.
} 
\end{figure}

Since the factorization property is predicted to hold only in the
$\beta-$relaxation regime, it is necessary that correlators show a
well developed plateau. From Fig.~\ref{fig_fsqzt_qdep} we recognize
that at $T=0.55$ this condition is not fulfilled for large values
of $z$ and therefore it is not possible to check whether or not the
factorization property holds. However, since these correlators are
very similar to the one in the bulk, and it has been shown that for
the bulk system the factorization property holds~\cite{gleim99}, we
can assume that it does so also for the confined system at large $z$,
{\it if} the temperature is sufficiently low.  For small values of $z$
it is instead possible to make a direct check on the validity of the
factorization property by calculating the function $R_x(t)$. The results
are presented in Fig.~\ref{fig_fsqzt.factorization} where we show the
time dependence of the function $R_x(t)$ for different values of $z$
(spaced by $\Delta z= 0.075$) and three different wave-vectors (panels
(a)-(c)). The times used were $t'=7.0$ and $t''=3000$. We recognize that
in the $\beta-$regime the correlators for the different values of $z$
fall nicely onto a master curve, thus showing that the factorization
property holds. Note that the fact that these curves fan out at short
and long times shows that the existence of such a master curve is by no
means a trivial matter.

The existence of the master curves shown in
Figs.~\ref{fig_fsqzt.factorization}a-c demonstrates that the shape of
the correlators is independent of $z$, as predicted by MCT. However,
this shape should also be independent of $q$. In order to check this
prediction we have calculated for several values of $q$ a mean master
curve by averaging the correlator $F^p_{\rm s}(q,z,t)$ over $0 <
z < 0.75$. The time dependence of these master curves for the different
wave-vectors are shown in Fig.~\ref{fig_fsqzt.factorization}d. From this
figure we see that the shape of the master curve does depend on $q$,
i.e. the factorization property does not hold. It might of course be
that the factorization property does hold at even lower temperatures,
but due to the finite size effects discussed in Sec.~\ref{sec3f} this
can presently not be checked.

\subsection{Structural relaxation times}
\label{sec3c}

\begin{figure}
\includegraphics[origin= 0 144, width=80mm]{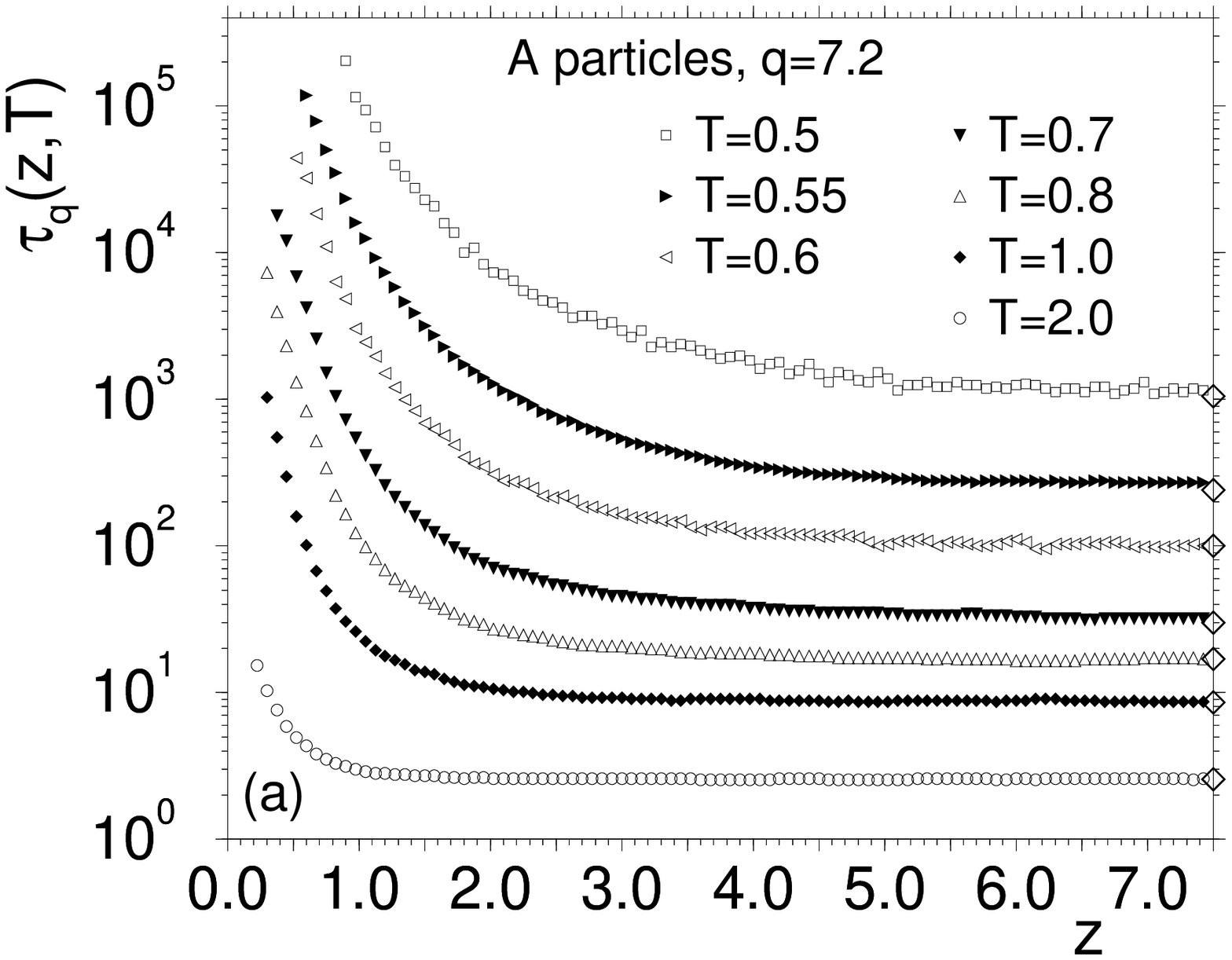}
\includegraphics[origin= 0 72, width=81mm]{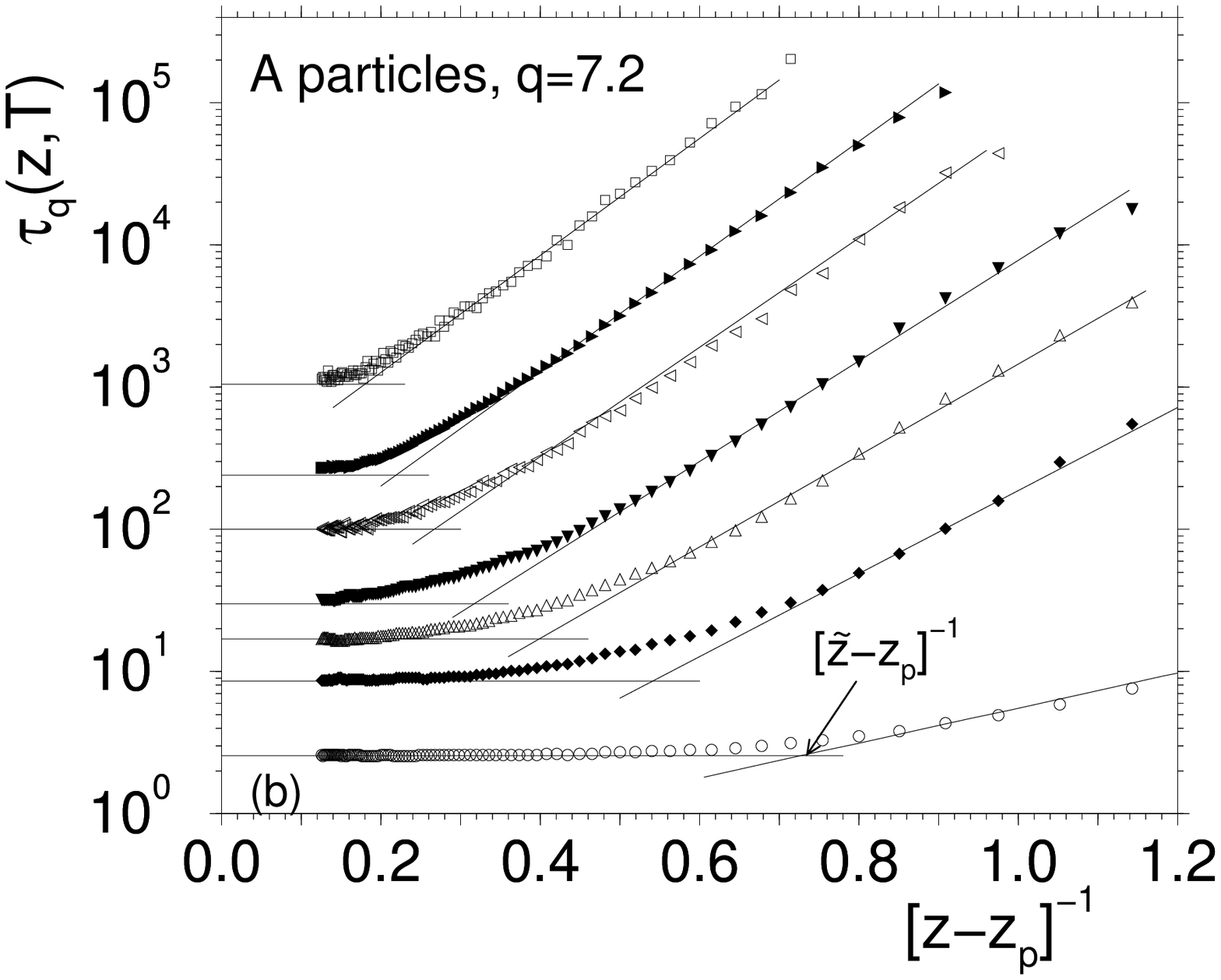}
\includegraphics[origin= 0 0, width=81mm]{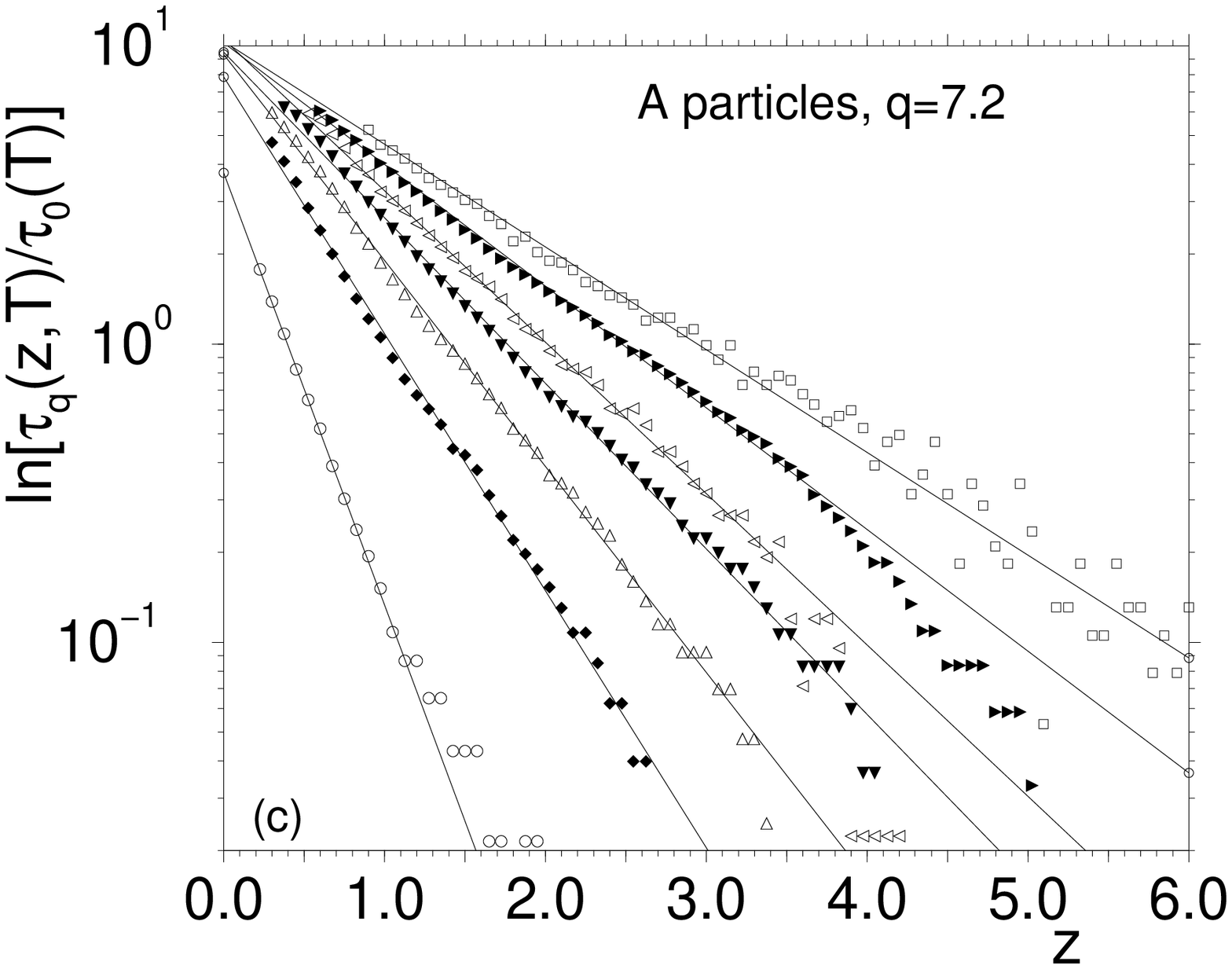}
\caption{\label{fig_tau_z}
$z$-dependence of relaxation times $\tau_q(z)$ from $F^p_{\rm
s}(q,z,\tau_q)=1/e$ at different temperatures (see legend in part (a)).
(a) Log-lin plot of the data. The open diamonds at $z=7.5$ are the relaxation times
in the bulk; (b) Reduced plot of $\tau_q$ versus $[z-z_{\rm p}]^{-1}$
with $z_{\rm p}=-0.5$. The horizontal lines are the bulk values of $\tau_q$. (c)
Plot of $\log [ \tau_q(z)/\tau_0 ]$ versus $z$ with $\tau_0$ as a
temperature dependent parameter. See main text for details.
}
\end{figure}

We will now discuss in more detail the spatial dependence of
structural relaxation times in the liquid. In Fig.~\ref{fig_tau_z}a
we plot $\tau_q(z)$, defined by $F^p_{\rm s}(q,z,\tau_q)=1/e$, for
all temperatures investigated. From this graph we recognize that the
thickness of the film is sufficiently large since at all temperatures
the relaxation times become independent of $z$ and are very close to the
value in the bulk (open diamonds at $z=7.5$). At the highest temperature
bulk behavior is realized already for $z \approx 2.0$. With decreasing
temperature the influence of the wall reaches further into the system,
i.e., the region which is affected by the wall grows with decreasing
temperature and at the lowest temperature considered it extends up
to $z\approx 5.0$. From the plot it also becomes evident that if the
temperature would be decreased somewhat more, the relaxation times in the
middle of the film would no longer coincide with the ones of the bulk,
i.e. that $\tau_q(z)$ becomes a function of $D$.

Our main goal is now to extract a characteristic dynamic length scale
from the data. Since there are no reliable theoretical concepts on how
this should be done, we are forced to use just an empirical description
of our data which includes such a length scale. In simulations of
the same Lennard-Jones liquid within a narrow cylindrical {\it pore}
\cite{scheidler00} we have been able to find such a description, at least
for the particles close to the walls, i.e. for small $z$. The proposed
Ansatz had the form

\begin{equation}
\tau_q(z)=f_q(T) \exp [ \Delta_q/(z-z_{\rm p}) ] \, ,
\label{eq15}
\end{equation}

\noindent
where $z_{\rm p}$ is a fit parameter that gives some sort of ``penetration
depth'' of the particles into the wall~\cite{scheidler00}. That this
functional form is indeed able to describe the data very well is shown in
Fig.~\ref{fig_tau_z}b. The solid inclined lines are fits with the function
given by Eq.~(\ref{eq15}) and we have chosen $z_{\rm p}=-0.5$. Since
the slope of these lines are the parameter $\Delta_q$, we see that
this quantity is increasing with decreasing temperature
(from 2.8 at $T=2.0$ to 7.5 at $T=0.5$). We note that this result
is in contrast to the one found for the narrow pore in that there
we found that the fit works well even if $\Delta_q$ and $z_{\rm p}$
are kept constant, i.e. independent of $T$~\cite{scheidler00}. The
reason for this difference might be that in Ref.~\cite{scheidler00}
the temperature at which the wall was frozen was always the same, in
contrast to the present case where $T_{\rm W}$ is set equal to $T$, see
Sec.~\ref{sec2}. (We remark that it is also possible to obtain good fits
to the data by keeping $\Delta_q$ fixed to 7.53 and to make $z_{\rm p}$
depend on temperature~\cite{scheidler_diss}.)

We now extract a characteristic length scale for the influence of the
wall by determining the length at which the fit with Eq.~(\ref{eq15})
crosses the bulk value (horizontal lines in Fig.~\ref{fig_tau_z}b). In the
following we will denote this so determined length scale by $\tilde{z}$
and in the next subsection we will discuss its $T-$dependence. (We remark
that the values of $\tilde{z}$ depend only weakly on the way one does
the fit to the data with Eq.~(\ref{eq15}), i.e.  whether $\Delta_q$
or whether $z_{\rm p}$ is kept constant.)

Since Eq.~(\ref{eq15}) is only capable to describe the data close to
the wall we looked for alternative possibilities to parameterize the
relaxation times. If one thinks of a characteristic length scale $\xi_0$
for a decay process, a natural Ansatz is the exponential form $\exp
[ -z/\xi_0]$. However, an analysis of the difference of $\tau_q(z)$
to the bulk value showed that the increase is much stronger than purely
exponential, and therefore we made an exponential Ansatz for the logarithm
of the ratio between $\tau_q(z)$ and a reference value $\tau_0$:

\begin{equation}
\ln \left[ \frac{\tau_q(z)}{\tau_0} \right]
= A(T) \cdot \exp \left[ - \frac{z}{\xi_0(T)} \right] \, .
\label{eq16}
\end{equation}

\noindent
In an ideal situation $\tau_0$ should of course correspond to
the bulk value since for large distances from the wall we have
$\tau_q(z)=\tau_q({\rm bulk})$. However, remaining small density
differences within the simulation as well as finite size effects (see
section \ref{sec3f}) lead to small deviation from this theoretical
value. Hence it was possible to improve the fit, in particular close to
the surface where the $z$-dependence is most pronounced, by extracting
$\tau_0$ from an extrapolation of $\tau_q(z)$ to large $z$ or by using
$\tau_0$ as a free parameter. This was done to obtain the data shown
in Fig.~\ref{fig_tau_z}c, where we plot in a logarithmic way the left
hand side of Eq.(\ref{eq16}) as a function of $z$. (We mention that the
values for $\tau_0$ differ only slightly from the bulk values.)  We see
that the data are indeed compatible with straight lines, and this for
all distances, thus validating the Ansatz.  The negative of the slope
of the straight lines is the inverse of $\xi_0$ and from the fact that
this slope decreases with decreasing temperature we see that $\xi_0$
becomes larger if $T$ is lowered. In the next section we will discuss
the $T-$dependence of $\xi_0$ in more detail. Finally we mention that
the length $\xi_0$ decreases slightly with increasing wave-number $q$,
but that this effect is only weak~\cite{scheidler_diss}.

\begin{figure}
\includegraphics[width=85mm]{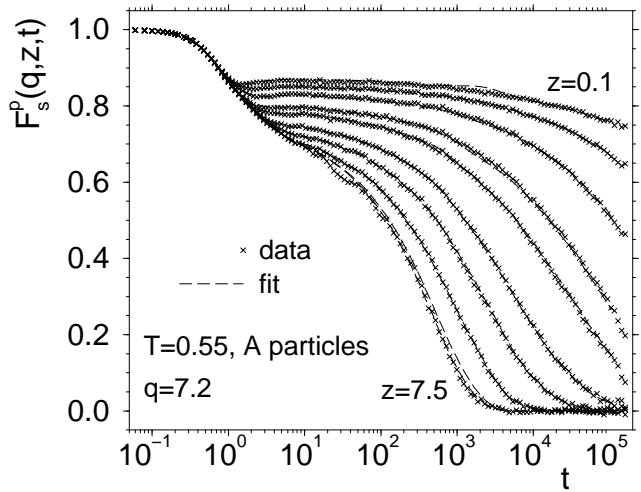}
\caption{\label{fig_fsqzt_parisifit}
Fit of the simulation data for $F^p_{\rm s}(q,z,t)$ for A particles at
$q=7.2$ and $T=0.55$ according to Eq.~(\ref{eq17}) for several values of $z$
($z=0.1,$ 0.3, 0.5, 0.7, 1.0, 1.5, 2.0, 3.0, and 7.5).}
\end{figure}

In Ref.~\cite{scheidler02b} we attempted not only to describe the
$z-$dependence of the relaxation times but the whole time and space
dependence of the scattering function $F^p_{\rm s}(q,z,t)$. The Ansatz used in
that study had the following form:

\begin{equation}
F^p_{\rm s}(q,z,t)=F_{\rm s}^{\rm bulk}(q,t)+
   a(t)\exp \left[ - \left( z/\xi(t) \right)^{\beta(t)} \right] \, .
\label{eq17}
\end{equation}

\noindent
Here $F_{\rm s}^{\rm bulk}(q,t)$ is the intermediate scattering
function in the bulk and thus the second term of the left hand side of
Eq.~(\ref{eq17}) describes the deviation of $F^p_{\rm s}(q,z,t)$
from this correlator. That this type of Ansatz is able to give a
very good description of the time dependent data is demonstrated in
Fig.~\ref{fig_fsqzt_parisifit} where we show $F^p_{\rm s}(q,z,t)$
for different values of $z$ (symbols) and the corresponding fit (dashed
lines. (We mention that if one plots the data versus $z$ for different
times $t$ the agreement between data and fit is very good as well (see
Ref.~\cite{scheidler02b}).)

\begin{figure}
\includegraphics[width=80mm]{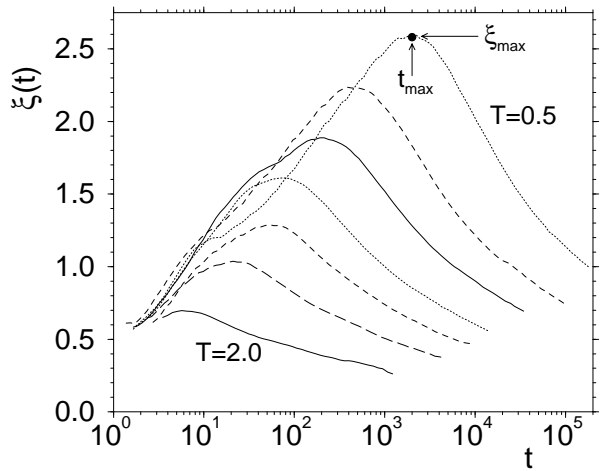}
\caption{\label{fig_fsqzt_parisifitpar}
Time and temperature dependence of the fit parameter $\xi(t)$ from
Eq.~(\ref{eq17}) at different temperatures ($T=2.0,$ 1.0, 0.8, 0.7, 0.6, 0.55,
and $0.5$).}
\end{figure}

The time and $T-$dependence of the stretching parameter $\beta(t)$ is
discussed in Ref.~\cite{scheidler02b}. Here we focus on the dependence
of the length scale $\xi(t)$ on $t$ and $T$. This dependence is shown in
Fig.~\ref{fig_fsqzt_parisifitpar} from which we recognize that $\xi(t)$
is a smooth function of time with a local maximum $\xi_{\rm max}$ at
$t=t_{\rm max}$. The location of this maximum shifts to larger times if
$T$ is decreased and also its value increases.

Thus we conclude that there exists a characteristic time scale at which
the influence of the wall extends farthest into the liquid, i.e. for which
the system is relatively stiff. This time scale corresponds roughly to the
time of the $\alpha$-relaxation in the bulk system~\cite{scheidler_diss},
a result which is plausible as can be seen as follows: It is well
known that for simple liquids the $\alpha-$relaxation is largest for
wave-vectors at the peak of the static structure factor (see, e.g.,
data in Ref.~\cite{kob95b}), i.e. the length scale that corresponds
to the typical distance between the particles. Since it is this time
scale which is really relevant for the structural relaxation, since
it corresponds to the typical time needed for the particles to leave
their cage, it can be expected that any external perturbation, such as
the presence of the wall, will affect the dynamics at long distances on
this time scale which is comparable with this relaxation time.

Finally we remark that the qualitative behavior as well as the
height of the maximum in $\xi(t)$ does not change significantly
if one fits the data with an exponential function, i.e. $\beta(t)
\equiv 1$ in Eq.~(\ref{eq17}), (see the subsequent section and
Fig.~\ref{fig_dynamic_lengthscales}), although we note that such a fit is
not able to describe the data very well~\cite{scheidler_diss}.

\subsection{Dynamic length scales}
\label{sec3d}

\begin{figure}
\includegraphics[width=80mm]{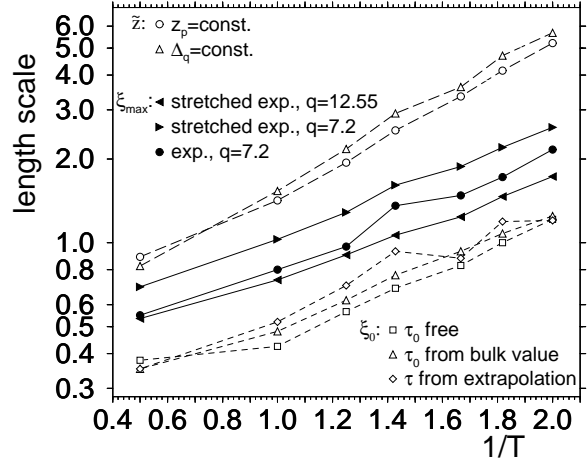}
\caption{\label{fig_dynamic_lengthscales}
Comparison of different dynamic length scales as extracted from the
spatial dependence of scattering functions $F^p_{\rm s}(q,z,t)$ and
structural relaxation times: $\tilde{z}$ from fits with Eq.~(\ref{eq15}),
$\xi_0$ from fits with Eq.~(\ref{eq16}) and $\xi_{\rm max}$ from fits
with Eq.~(\ref{eq17}) (see also legend and text). 
}
\end{figure}

From the results described so far we were able to extract three
different dynamic length scales: $\tilde{z}$ from Eq.~(\ref{eq15}),
$\xi_0$ from Eq.~(\ref{eq16}), and $\xi_{\rm max}$ from the Ansatz
given by Eq.~(\ref{eq17}). We now discuss how these different
length scales depend on temperature. This $T-$dependence is shown in
Fig.~\ref{fig_dynamic_lengthscales} where we have plotted $\tilde{z}$,
$\xi_0$, and $\xi_{\rm max}$ as a function of inverse temperature. First
of all we recognize that the different definition of length scale give
basically all the same result and that thus the precise definition
is not relevant. Furthermore we see that it is also not crucial
how exactly one defines a length scale: For the case of $\tilde{z}$ it
does not manner whether one keeps $\Delta_q$ or $z_{\rm p}$ independent
of $T$ (see discussion of Eq.~(\ref{eq15})); for the case of $\xi_0$
it does not matter whether the relaxation time $\tau_0$ is the bulk
value, the value obtained from extrapolating $\tau_q(z)$ to large $z$,
or whether it is just a fit parameter; for the case of $\xi_{\rm max}$
it is irrelevant at which wave-vector one uses the Ansatz (\ref{eq17})
or whether $\beta(t)$ is set to unity or not.

Note that we have plotted all data in an Arrhenius plot, i.e.
logarithmically versus inverse temperature. Although to our knowledge
there is no theoretical reason why one should plot the data in this
manner, the graph shows that this presentation does rectify the data.
Thus we conclude that the length increases like $\exp(B/T)$ and
that the value of $B\approx 1.2$ is independent of the length scale.
In Ref.~\cite{scheidler02a} we have discussed the temperature dependence
in more detail and made also a comparison with other characteristic
length scales in the liquid. The conclusion is that basically all
relevant length scales are very similar in size and show a quite
similar $T-$dependence. Finally we remark that this $T-$dependence is
significantly larger than the one that one obtains from the one of static
quantities like, e.g., the envelope of the radial distribution
function~\cite{scheidler02a}.

\subsection{Collective quantities}
\label{sec3e}

To get a better understanding of the collective dynamics close to the
surface we investigate the intermediate scattering function $F(q,z,t)$
for all (A and B) particles which is given by Eq.~(\ref{eq10}).
In this quantity the initial position of a particle is correlated
with the position of any other particle at a later time $t$. The main
consequence for the behavior close to the surface is the following. Since
the configuration of the wall particles is frozen there exist a static
landscape in the potential energy and particles of the liquid prefer to
occupy the local minima in this landscape. Therefore it can be expected
that as soon as a liquid particle close to the surface leaves its position
this vacancy will be filled by another particle. This implies that the
collective correlation function close to the wall does not decay to zero
even at long times.

\begin{figure}
\includegraphics[width=80mm]{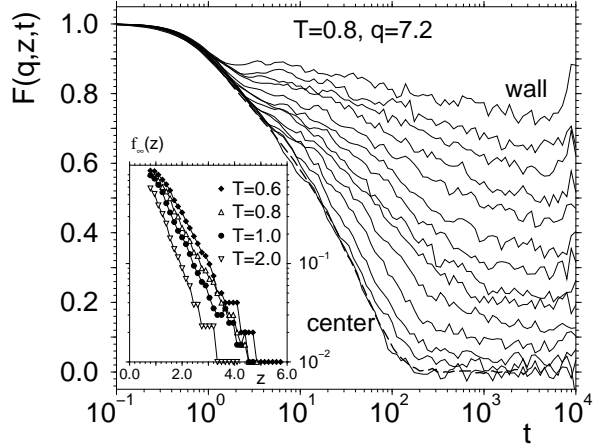}
\caption{\label{fig_fqzt} 
Normalized coherent intermediate scattering function $F(q,z,t)$ for $q=7.2$ for both types
of particles at $T=0.8$.  The thin curves correspond to $z=0.3, 0.45,
0.6, \ldots , 1.35, 1.5, 1.8, 2.25, 2.85, 3.9, 7.5$. The bold dashed line is the data
for the bulk. Inset: $z-$dependence of the height of the plateau $f_{\infty}(z)$
for different temperatures.}
\end{figure}

That this expectation is indeed met is shown in Fig.~\ref{fig_fqzt}
where we plot the time dependence of $F(q,z,t)$ as a function of
temperature. Note that, in order to improve the statistics of the data,
we do not distinguish between A and B particles and in addition we
have made a spherical average. We mention, however, that the data in
which we averaged only over wave-vectors parallel to the surfaces look
qualitatively quite similar~\cite{scheidler_diss}.

Also included in the figure is the corresponding correlator for the bulk
system (bold dashed line) and we recognize that this function coincides
with $F(q,z,t)$ for large values of $z$. Furthermore we see that the
height of the plateau in $F(q,z,t)$ at long times increases rapidly
with decreasing $z$ and that for the smallest $z$ we do no longer
find the two-step relaxation usually seen in supercooled liquids. The
$z-$dependence of the height of this plateau, $f_{\infty}(z)$, is shown
in the Inset of the figure (in a log-lin plot). We see that it shows
basically an exponential decay, although within the noise of the data it
is difficult to exclude other functional forms. The slope of the straight
lines in the Inset gives a {\it static} length scale over which the liquid
starts to become uncorrelated with the ``pinning-field'' generated by
the wall. Thus it can be expected that this length scale is comparable to
the one found in other static two-point correlation functions, such as,
e.g. the decay of the correlation in the radial distribution function. In
agreement with this expectation we see that this length scale is indeed on
the order or 2-3 and that it shows only a weak dependence on temperature.
In particular this $T-$dependence is significantly weaker than the one
found for the dynamical length scales discussed above.

Due to the presence of the plateau at long times in $F(q,z,t)$, it is
difficult to extract a relaxation time from this function. In an attempt
to define such a time scale, we have subtracted the long time limit of
$F(q,z,t)$ and normalized the function by $1-f_{\infty}(z)$. For the
so obtained resulting normalized correlator it is possible to define
a relaxation time. However, we have found that this time scale depends
only relatively weakly on $z$, and in particularly much weaker than the
relaxation time for $F_s^p(q,t,z)$, and therefore this approach was no
longer pursued~\cite{scheidler_diss}.

\subsection{Films with a finite thickness}
\label{sec3f}

Until now we have considered only values of $D$ and $T$ such that the
system under investigation is sufficiently large to guarantee that in the
center it has the same dynamics as in the bulk. Thus strictly speaking we
have so far only investigated the dynamics of a liquid in the vicinity
of {\it one} wall and the fact that the overall geometry is a film
was completely irrelevant. However, from Fig.~\ref{fig_tau_z} we have
concluded that with decreasing temperature the dynamical length scales
for cooperativity are growing and that the influence of the wall reaches
further into the system. Therefore it is obvious that for a system of
a given thickness also the particles in the center will be influenced by the
walls, if the temperature is sufficiently low. Hence it can be expected
that the dynamics should be slowed down even more and the goal of this
subsection is to investigate this effect in more detail.

\begin{figure}
\includegraphics[width=80mm]{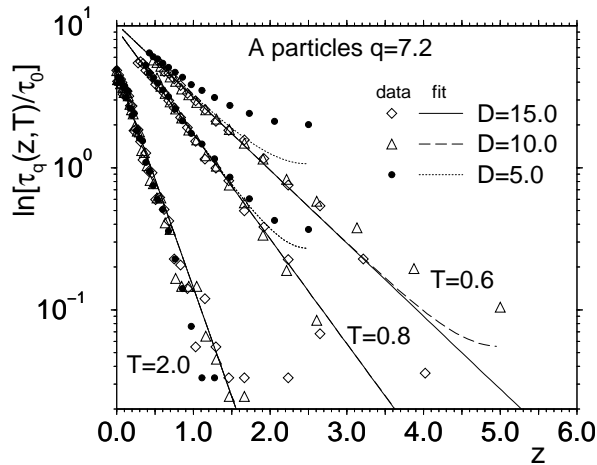}
\caption{\label{fig_tau_z_finitesize}
Local relaxation times in films with different thickness ($D=15.0,$ 10.0, and
5.0).  Plot of the left hand side of Eq.~(\ref{eq16}) versus distance
to the nearest wall (symbols). The lines correspond to the Ansatz given
by Eq.~(\ref{eq16}) with parameter that have been determined in systems
with $D=15.0$.}
\end{figure}

Instead of lowering the temperature we investigate films that have
a smaller thickness ($D=10.0$ and $D=5.0$) for temperatures down
to $T=0.6$. We have repeated the procedure described in the previous
subsection for $D=15.0$ in order to obtain the $z-$dependence of the
relaxation times of the incoherent intermediate scattering function
and we present this dependence in Fig.~\ref{fig_tau_z_finitesize}.
At high temperatures, where each particle is influence by at most {\it
one} wall, the local dynamics is only a function of its distance to the
wall. This can be see by the data for $T=2.0$ since the curves for the
different values of $D$ fall on top of each other and can, e.g.,
be described by the exponential Ansatz  of Eq.~(\ref{eq16}), solid line.
If we decrease temperature, $T=0.8$, the data for the thinnest film starts
to deviate from the master curve because the particles in the center
already feel the presence of both walls. At even lower temperatures,
$T=0.6$, this effect becomes more pronounced and also the data for
$D=10.0$ starts to deviate from the one of $D=15.0$. Note that since
at this $T$ the data for $D=10.0$ shows strong deviations from the one
for $D=15.0$ at $z=5.0$, it follows that in a film of thickness $D=5.0$
the particles close to a wall are also influenced by the presence of
the {\it opposite} wall.

The question is now if this enhancement of the slowing down is just a {\it
linear} effect, i.e. it is given by the sum of the contributions
from both walls, or if the situation is more complicated. Under the
assumption that the local dynamics at a distance $z$ from a rough surface
is given by Eq.~(\ref{eq16}) a reasonable generalization is given by the Ansatz

\begin{equation}
\ln \left[ \frac{\tau_q(z)}{\tau_0} \right]
= A(T) \cdot \left\lbrace
    \exp \left[ - \frac{z}{\xi_0(T)} \right] +
    \exp \left[ - \frac{D-z}{\xi_0(T)} \right]
	     \right\rbrace \, .
\label{18}
\end{equation}

Using the parameters $A(T)$ and $\xi_0(T)$ as determined from the fits
to the data for $D=15.0$, see Sec.~\ref{sec3c}, we thus can use this
expression to calculate the $z-$dependence of $\tau_q$ for the smaller
films. The resulting predictions for $D=10.0$ and $D=5.0$ are included
in Fig.~\ref{fig_tau_z_finitesize} as well (dashed and dotted lines). We
see that although the predicted curves look qualitatively similar to
the data for these films, they severely underestimate the relaxation
times at large values of $z$. Qualitatively the same result is obtained
if we use Eqs.~(\ref{eq15}) or (\ref{eq17}) to predict $\tau_q(z)$ for
thin films~\cite{scheidler_diss}. Hence we conclude that the presence
of two walls affects the relaxation dynamics of the liquid particles in
a non-additive manner, i.e. that the reason for the slowing down must be
non-linear. This result is not that surprising since mode-coupling theory,
a very successful theoretical framework to describe the slowing down of
the dynamics of glass-forming systems, predicts that this slowing down
is related to a non-linear feedback mechanism~\cite{mct}. According to
MCT the relaxation dynamics of a particle is strongly coupled to the
motion of its neighboring particles that form the cage of the tagged
particle. Since this tagged particle itself is part of the cage of other
particles, a decrease in $T$, i.e. a stronger coupling of the motion,
leads to a strong (i.e. non-linear) stiffening of the cage (feedback
effect). In our case the particles that form the wall are completely
frozen and thus they do no longer participate to help to relax the fluid
particles in their neighborhood. Thus from a qualitative point of view
the lack of fluctuations of the cage of the particles near the walls
explains the slowing down of their dynamics and the feedback effect
rationalizes why the presence of two walls affects the system stronger
than just the linear sum of two individual walls. Finally we mention that
this non-linear feedback effect should affect the relaxation dynamics
even more in the case of systems in which the confinement is given by
a tubular geometries or in cavities.

\subsection{Mean relaxation dynamics}
\label{sec3g}

Having characterized in great detail the local dynamics and understood how
the structural relaxation depends on the distance from the wall, we now
can investigate the dynamic properties of the whole system, i.e. if one
averages the correlators over $z$. These averaged correlation functions
are exactly the quantities that are accessible in most experiments. In
contrast to experiments we are, however, in the favorable position to
be able to relate these averaged quantities to the time dependence of
the local ones.

\begin{figure}
\includegraphics[width=80mm]{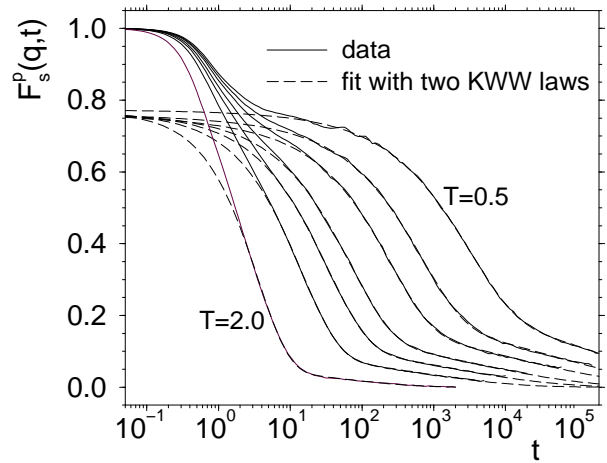}
\caption{\label{fig_fsqt.film} 
Self part $F^p_{\rm s}(t)$ of the intermediate scattering
function for A particles in a film with $D=15.0$ for $q=7.2$ and
temperatures $T=2.0, 1.0, 0.8, 0.7, 0.6, 0.55$, and $0.5$ (solid lines).
The dashed lines are fits to the data with a superposition of two
stretched exponential laws.}
\end{figure}

In Fig.~\ref{fig_fsqt.film} we show the time dependence of $F^p_{\rm
s}(q,t)$, i.e. the incoherent intermediate scattering function for the A
particles averaged over the whole film (solid lines). (The wave-vector
is again parallel to the walls.) Although at a first glance these
correlators look quite similar to the ones in the bulk~\cite{kob95b}, a
closer inspection shows that there are important differences.  Already at
$T=2.0$ one sees that after a rapid decay to a small value around 0.05,
the correlator shows a tail that extends to relatively long times. At low
temperatures the function shows at intermediate times the usual plateau
that is related to the cage effect but at longer times the mentioned
tail starts now to become very pronounced.

The reason for this behavior follows directly from the results of the
local correlation functions. Since the typical time scales for the
structural relaxation as a function of $z$ extend over several decades
in time, the average over all those curves should be very stretched
(see also Fig.~\ref{fig_fsqzt}). For $T=2.0$ particles with
$z>2.0$, i.e. more than 70\% of the particles, behave more or less
bulk-like and contribute to the fast decay seen in Fig.~\ref{fig_fsqt.film}
at short times.  The correlators for particles at the surface decay
much slower and give therefore rise to the long time tail for $t>10$. With
decreasing temperature the number of particles with a relaxation that is
significantly slower than the one in the bulk becomes larger and therefore
the amplitude of the long time tail increases. As a consequence the shape
of the correlators change with temperature, i.e. the time-temperature
superposition principle is not valid, in contrast to the case of the
bulk system~\cite{kob95b}. Therefore the definition of characteristic
relaxation times for the whole decay is not really possible.

Of course one can calculate the time dependence of $F^p_{\rm
s}(q,t)$ from the knowledge of the $z-$dependent correlators
which we are able to describe by a set of KWW functions (see
Fig.~\ref{fig_fsqzt_superpos}). On the other hand, in experiments this
information is usually not available and therefore it is interesting
to check how one can characterize the averaged data in a different way.
We have found that for all temperatures considered, the $\alpha-$relaxation
of the average correlator can be described very well by the sum of two
KWW-laws (having different amplitudes): One that is for the early part
of the relaxation and the second one for the mentioned tail. That this
type of Ansatz does indeed give a very good description of the data is
shown in Fig.~\ref{fig_fsqt.film} where we have included the results of
such fits as well (dashed lines).

\begin{figure}
\includegraphics[width=78mm]{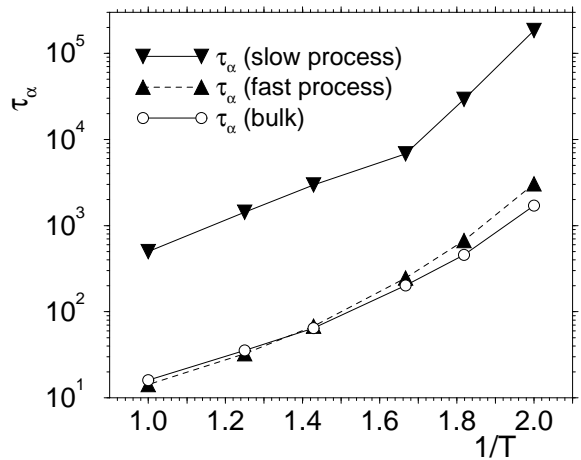}
\caption{\label{fig_fsqt_fitpar}
Temperature dependence of the relaxation times of the two processes
extracted from the self part of the intermediate scattering function
averaged over all A particles in the system (filled triangles).
The open circles are the $\alpha-$relaxation time for the bulk.}
\end{figure}

In Fig.~\ref{fig_fsqt_fitpar} we show the temperature dependence of
the relaxation times, obtained from the KWW fits, for the fast and
slow process. We see that within the accuracy of the data i) these
$T-$dependencies are very similar, although the time scale for the
slow process is around 40 times larger, and ii) that the fast process
shows basically the same $T-$dependence as the bulk system. For the
fast process the values for the KWW-parameter $\beta$ is around 0.8
and depends only weakly on $T$. In contrast to this we find for the
slow process that $\beta$ is only around 0.25-0.3, i.e. a very small
value~\cite{scheidler_diss}. The reason for such a strong stretching,
which is basically independent of $T$, is that the relaxation dynamics
is extremely heterogeneous, i.e. that close to the wall the typical time
scales differ by orders of magnitude (see Fig.~\ref{fig_fsqzt}).

Thus we find that the long time tail is basically given by the second
process which has a clearly separated time scale with a smaller
amplitude. Note that this conclusion comes from our data in the {\it
time} domain. However, it is also possible to make a time-Fourier
transform of our data and to calculate the dynamical susceptibility
$\chi(\omega)$. This has been done in Ref.~\cite{scheidler02a} where it
was shown that these two processes can also be seen in the frequency
domain in that one finds that the imaginary part of $\chi(\omega)$
seems to be indeed the sum of two peaks. Such a double peak structure
is often found in real spectroscopy experiments where one observes a
second peak at frequencies well below the peak corresponding to the
$\alpha-$relaxation of the bulk system. Thus from such {\it averaged}
data one could easily come to the conclusion that there are indeed two
distinct processes, e.g. one corresponding to particles in the center
and the other to a layer of slow particles at the surface and which give
rise to the second process. However, the analysis of the $z-$dependent
correlators presented here shows that such a conclusion might be wrong.

\begin{figure}
\includegraphics[width=80mm]{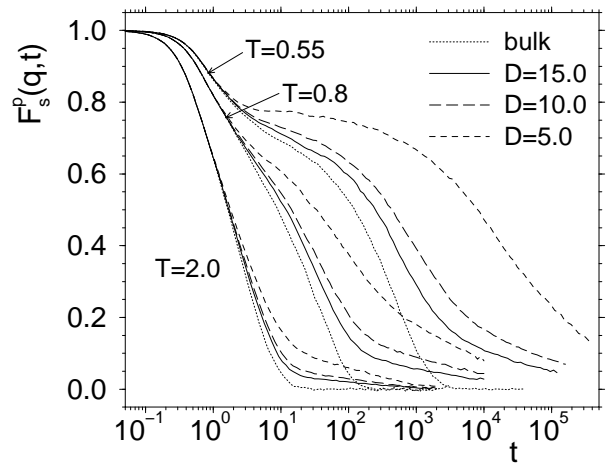}
\caption{\label{fig_fsqt.film.size}
Time dependence of the intermediate scattering function $F^p_{\rm s}(q,t)$
for all A particles in films with different thickness ($D=15.0$, 10.0,
5.0) for $q=7.2$ and different temperatures ($T=2.0$, 0.8, 0.55). The
dotted curves correspond to the bulk data.  
} 
\end{figure}

Finally we discuss how the averaged time correlation function depends
on the thickness of the film. In Fig.~\ref{fig_fsqt.film.size} we show
$F^p_{\rm s}(q,t)$ for different temperatures and three values of $D$
as well as for the bulk system. We see that in these averaged curves one
can notice already at the highest temperature a significant deviation of
the correlators for the confined systems from the one of the bulk. This
is due to the presence of the particles that are slowed down near {\it
one} of the walls. (Recall that in Fig.~\ref{fig_tau_z_finitesize} we
have seen that at this $T$ none of the particles of even the smallest
system are affected by {\it both} walls.) Thus the amplitude of the
tail at long times is given only by twice the fraction of the particles
that are strongly affected by one wall. (Note that the value of this
fraction depends on $D$.) With decreasing $T$ the fraction of particles
affected by one wall increases, due to the growing dynamical length
scale and hence the curves for the confined system decay significantly
slower than the bulk curves. For even lower temperatures we have the
additional effect that the particles close to the center are affected in
a non-linear way by the presence of {\it both} walls, which leads to an
additional slowing down. Thus we see that the $T-$ and $D-$dependence
of the relaxation function is quite complex since there are three main
causes: i) The general slowing down of the dynamics with decreasing
temperature; ii) The presence of a growing length scale over which the
slow dynamics close to the wall affects the relaxation dynamics; iii)
The non-linear effects occurring if a particle is influenced by more
than one wall. The sum of these causes has the effect that the shape
of the correlation function depend in a strong and highly non-trivial
way on $T$ and $D$. This dependence also hampers a sensible definition
of a relaxation time which is the reason why we do not discuss the $T-$
and $D-$dependence of it.

\section{Summary and conclusion}
\label{sec4}

In this paper we have presented the results of a molecular dynamics
computer simulation of a binary Lennard-Jones liquid confined between two
rough surfaces. The walls are given by a frozen amorphous configuration of
the same liquid and as we have shown here that the static properties of the confined
fluid are identical to the ones of the bulk system. We have found that the
relaxation dynamics of particles close to the interface are slowed down
by orders of magnitude with respect to the one in the bulk. Due to the
cooperativity of the particle motion within a liquid, also particles at
some distance from the wall are influenced by the presence of the wall.
Therefore the properties of the structural relaxation become a function
of distance $z$ to the wall. All these properties are a smooth function
of $z$, although some of them show a very strong $z-$dependence which
results that the dynamics of the system is very heterogeneous.

The time dependence of the incoherent intermediate scattering functions
$F_{\rm s}(q,z,t)$ is described well by a KWW-function with a stretching
parameter $\beta$ and a relaxation time $\tilde{\tau}$ that depends
strongly on $z$. The $z-$dependence of the relaxation times can be
described well by various phenomenological laws. From these laws it
is possible to extract various dynamical length scales over which the
dynamics of the system is influenced by the walls. 

These length scales are of purely dynamical origin and show a
$T-$dependence which is compatible with an Arrhenius law with
an activation energy that is independent of the details of the
definition of the length scale. Note that this $T-$dependence
is in contrast to other dynamical length scales that have been
discussed in the literature before, such as, e.g., the dynamical
heterogeneities~\cite{donati99a,franz99,franz00} for which a divergence at
a temperature close to $T_c$ of mode-coupling theory has been found. We
have tested whether or not the $T-$dependence of our length scales are
compatible with a divergence at a finite temperature and have found, see
Ref.~\cite{scheidler_diss}, that in principle such a scenario is possible
at $T_c$ or at $T_K$, the Kauzmann temperature of the system. However,
the noise in the data is unfortunately too large to make strong statements
on this point.

As long as the dynamical length scales are significantly smaller than the
system size, the discussed results are just an effect due to the interface,
i.e. the dynamics is only a function of distance to the surface
and independent of the film thickness. For a film with given thickness
there exists however a threshold temperature below of which particles are
influenced by both interfaces since the size of the CRR's are larger than
$D/2$. In that case we find that the slowing down is much stronger than
one might expect from a simple linear superposition of the influence of
two single surfaces. This is evidence that the mechanism for the slowing
down of the particles is a strongly non-linear effect, such as proposed,
e.g., by mode-coupling theory.

Finally we have studied the time dependence of the correlators averaged
over the whole film.  These correlators are a quite complex function
of $T$ and $D$ since one has to distinguish between the normal slowing
down of the liquid with decreasing temperature (which is present in the
bulk system as well), the influence of one interface on the dynamics
with a dynamical length scale that increases with decreasing $T$, and
last not least with the mentioned non-linear effects. In view of these
different reasons for the slowing down of the average dynamics, it is
clear that it is rather difficult to come up with an expression for the
$D-$dependence of the glass transition temperature, since it is not even
obvious how the typical relaxation time of the system should be defined.

Despite this uncertainty it is clear that the glass transition
temperature, which could be defined as the temperature at which a
relaxation time or the viscosity attains a given predescribed value,
will increase with decreasing thickness of the film. However, we
point out that this qualitative result holds only true for the case
of rough walls as they have been considered in the present work.  In a
different study we have also investigated the relaxation dynamics of
a system with {\it smooth} walls where we have again made sure that
the structural properties of the confined liquid is not changed by
these interfaces~\cite{scheidler_diss,scheidler02a}. In that study we
have found that the dynamics close to the wall is {\it accelerated}
which in turn leads to a decrease of the glass transition temperature
with decreasing $D$. Hence we conclude that the relaxation dynamics is
strongly affected by the nature of the confining walls, in agreement
with experimental results. Thus it seems that it would be of interest
to avoid these confining walls altogether in order to disentangle the
influence of the wall on the dynamics and the effect of the finite
extension of the system. Within a simulation this is indeed possible
since one can investigate the relaxation dynamics of a given system for
different system size (using periodic boundary conditions in all three
directions). In the past such simulations have been done and it has been
found that at a given temperature the relaxation dynamics slows down with
decreasing temperature~\cite{buchner99,kim00}. Hence we conclude that intrinsically
confinement leads to a slower dynamics but that the presence of fluid-wall
interactions can completely mask this general trend.

Last not least we comment on the necessity for a more complete theoretical
understanding of the relaxation dynamics of confined systems. The
present simulations have shown that it is possible to have a slowing
down of the dynamics without a change of the structural properties
(and in Refs.~\cite{scheidler03,scheidler02a} it was shown that also
an acceleration can be observed). Hence we conclude that it is not
sufficient to know the structure in order to predict the dynamics. This
is in contrast to the situation of the bulk where it is indeed possible,
using mode-coupling theory, to predict the relaxation dynamics using
as input only structural quantities~\cite{mct,kob02}. For the moment
it is not clear to us how this theory has to be modified in order to
rationalize the results found in this work. Of course it is probably
possible to use some phenomenological theories in order to describe
these results~\cite{herminghaus01,long01,mccoy02} but having a theory
at hand that allows a full microscopic calculation would certainly be
much more preferable.

Acknowledgement: Part of this work was supported by the DFG under SFB 262,
by the European Community's Human Potential Programme under contract
HPRN-CT-2002-00307, DYGLAGEMEM, and the NIC in J\"ulich.


\end{document}